\documentclass[a4paper,twocolumn,showkeys,floatfix,aps,prd,amsmath,amssymb,nofootinbib,preprintnumbers,superscriptaddress]{revtex4-1}
\pdfoutput=1
\usepackage{times,amsfonts}
\usepackage{natbib}
\usepackage[english]{babel}
\usepackage[T1]{fontenc}
\usepackage[latin2]{inputenc}
\usepackage{graphicx}
\usepackage{color}
\usepackage{tabularx}
\usepackage{mathrsfs}

\bibpunct{(}{)}{;}{a}{}{,}
\usepackage{aas_macros}

\usepackage{attachfile2}

\usepackage{hyperref}
\usepackage[hyphenbreaks]{breakurl}
\usepackage{url}

\hypersetup{
colorlinks=true,  
urlcolor=blue,    
linkcolor=red,    
}

\makeatother

\newcommand{\Beginruledtabular}{\begin{ruledtabular}}
\newcommand{\Endruledtabular}{\end{ruledtabular}}

\newcommand{\stageOne}{{\it stage-one }}
\newcommand{\stageTwo}{{\it stage-two }}
\newcommand{\stageThree}{{\it stage-three }}
\newcommand{\fastTrack}{{\tt fast\_track}}
\newcommand{\coconut}{{\tt COCONUT}}
\newcommand{\Aoa}{{A_{\rm oa}}}
\newcommand{\Zoa}{{Z_{\rm oa}}}
\newcommand{\Model}[1]{M^{(#1)}}
\newcommand{\ModelA}[1]{M^{(#1)}_A}
\newcommand{\ModelZ}[1]{M^{(#1)}_Z}
\newcommand{\citeLewInPrep}{B.S. Lew, 2018, in preparation}
\newcommand{\RT}{32-m Toru\'n radio telescope}

\usepackage{ifthen}
\renewcommand{\eprint}[2][arXiv]{
\ifthenelse{\equal{#1}{ascl}}
{\href{http://www.ascl.net/#2}{\tt #1:#2}}
{\href{http://arxiv.org/abs/#2}{\tt #1:#2}}
}

\providecommand{\adsurl}[1]{[\href{#1}{\tt ADS}]}
\begin{document}

\title{Improving pointing of Toru\'n 32-m radio telescope:
effects of rail surface irregularities}

\author{Bartosz Lew} \email[]{bartosz.lew@umk.pl }
\affiliation{Centre for Astronomy, Faculty of Physics, Astronomy and Informatics, Nicolaus Copernicus University, Grudziadzka 5, 87-100 Torun, Poland}

\date{Oct 11, 2017}

\begin{abstract}
Over the last few years a number of software and hardware improvements
have been implemented to the 32-m Cassegrain radio telescope located
near Toru\'n. The 19-bit angle encoders have been upgraded to 29-bit
in azimuth and elevation axes.  The control system has been
substantially improved, in order to account for a number of
previously-neglected, astrometric effects that are relevant for
milli-degree pointing.  In the summer 2015, as a result of maintenance
works, the orientation of the secondary mirror has been slightly
altered, which resulted in worsening of the pointing precision, much
below the nominal telescope capabilities.

In preparation for observations at the highest available frequency of
30-GHz, we use One Centimeter Receiver Array (OCRA), to take the most
accurate pointing data ever collected with the telescope, and we
analyze it in order to improve the pointing precision.

We introduce a new generalized pointing model that, for the first
time, accounts for the rail irregularities, and we show that the
telescope can have root mean square pointing accuracy at the level
${<}8''$ and ${<}12''$ in azimuth and elevation respectively.
Finally, we discuss the implemented pointing improvements in the light
of effects that may influence their long-term stability.

\end{abstract}


\maketitle

\section{Introduction}
The Toru\'n 32-meter radio telescope is a classical Cassegrain
telescope with alt-azimuth, wheel-on-track mounting.  Located in central Europe,
and operated 24-h a day, it is one of the European VLBI
Network (EVN) nodes, capable of observing at frequencies from ${\sim}1$
GHz to ${\sim}30$ GHz in continuum and spectroscopic modes
at the selected bands.

Over the last several months a few important astrometric improvements
have been introduced into the control system.
More improvements were implemented over the period of last few years,
including weather dependent radio refraction and corrections due to
differences between the Coordinated Universal Time (UTC) and Universal
Time (UT1). The secondary mirror of the Cassegrain system by design
has five degrees of freedom: 2 for lateral translations in the focal
plane, 1 for translations along the optical axis, and 2 for rotations
about secondary focus. Given that the mirror drives become unreliable
over time in 2015 the mirror has been fixed.  As a result, its
orientation and gravitational sag have changed. Furthermore, the
19-bit angle encoders have been upgraded in azimuth and elevation axes
in 2013 and 2014 respectively, and now provide position information
with sub-arcsecond resolution.

For about ten years, since the initial fixation of the secondary
mirror in 2006, and since introducing the pointing model used for the
observations presented in this work, it has been known that the
residuals between the measured pointing corrections, and the
best-fitting pointing model, exhibit patterns that vary faster in
angular space than the model can accommodate.  However, only with the
advent of higher precision 22-GHz spectral pointing observations, and
the data acquired in the continuum with OCRA-p (One Centimeter
Receiver Array prototype) radiometer \citep{Browne2000}, and by using
an improved version of the control system, which provides position
readouts at higher frequency, and also by improving the extraction
algorithms for position corrections from cross-scan observations, in
the year 2015, it become possible to undoubtedly associate these high
frequency patterns to irregularities of the rail as the telescope
trolleys roll over the welding points.  All of these observations
require introducing a more general pointing model and its calibration
by means of the new pointing data.  In preparation for the observation
programs carried out at frequencies above 20~GHz
\citep{Lowe2007,Gawronski2010,Lancaster2011,Lew2015,Lew2016b, Peel2011, Szymczak2016}, we
carry out pointing campaigns aiming at improving the pointing
precision to one-tenth of the highest frequency beam,
i.e. ${\sim}0.002^\circ$.

In this work we introduce a generalized pointing model, implement it
into the control system and discuss the resulting improvements.
The assumed target precision, the order of one milli-degree, may also
require precise thermal control of the supporting structure since
seasonal, day to night and sun to shade temperature variations may
have significant effects on pointing capabilities of large-aperture
and/or millimeter-wave telescopes
\citep{vonHoerner1967,Baars1988,Bayley1994,Ukita1999,Prestage2004,Orfei2004,Greve2006,
Pisanu2010,Sun2014,Bolli2015}. Other effects such as strong wind can also
impact surface deformation and pointing at the level of several seconds
of arc \citep{Ukita2007,Ukita2008} and lead to significant loses of available
observing time \citep{Ries2011,Ries2012}.

It is known that the surface accuracy of the rail, its design and
component connections may also significantly impact pointing
capabilities of large telescopes \citep{Shi2014,Chen2016, Li2017,
Kaercher2004,Kong2014}. While controlling parameters such as
temperature or inclination of the structure may also be critical
for maintaining high pointing performance, in the present work on the
32-m Toru\'n radio telescope, we primarily focus on modeling the
previously-neglected effects associated with the telescope rail, and
how they limit pointing capabilities, while we leave discussing
thermal effects to a separate analysis.

The impact of the rail surface irregularity on the orientation
of the antenna, and consequently on the amplitude of the associated
position corrections can be accurately modeled using data obtained
from inclinometers installed on the structure of the telescope
(e.g. \citealt{Gawronski2000,Gawronski2006}).  In the absence of the
inclinometers, an alternative approach is possible, which is to
extract the position corrections directly from pointing observations.  

In section Sec.~\ref{sec:setting} we lay out the notation and basic
relations used throughout the rest of the paper.  In section
Sec.~\ref{sec:model} we describe pointing models used with
the {\RT}. Pointing observations and data processing are discussed in
Sec.~\ref{sec:observations} and Sec.~\ref{sec:data} respectively.  The
main results are presented in Sec.~\ref{sec:bestfit4e} and
Sec.~\ref{sec:model5} where we calibrate the extended pointing models
and present the impact of rail irregularities.  We discuss the results
in connection with independent measurements of the rail in
Sec.~\ref{sec:discussion} and then conclude in
Sec.~\ref{sec:conclusions}.

\section{Coordinates and reference frames}
\label{sec:setting}
When an observer demands to track a distant source at equatorial
coordinates $(\alpha_e,\delta_e)$ for a given epoch (e.g. J2000),
these coordinates are converted to $(\alpha_S, \delta_S)$ coordinates
of the current date, accounting for precession, nutation, and annual
aberration. The corresponding horizontal celestial coordinates
$(A_S,Z_S)$ of the date are calculated using standard spherical
trigonometry rotations for the actual local mean sidereal time.
These coordinates are referred to as ``SET'' coordinates and they
represent the actual, in-vacuum direction towards the source in
horizontal coordinates.

In order to track the source, the control system needs to minimize the
difference between the updated $(A_S,Z_S)$ direction and the ``TRUE''
(in-vacuum) direction of the selected telescope beam $(A_T,Z_T)$,
accounting for its offset $(B_A,B_Z)$ from the telescope optical axis. The
corresponding $(\alpha_T,\delta_T)$ are calculated and updated using
$(A_T, Z_T)$ for the actual time.  When tracking, the horizontal
coordinates of the optical axis are related to the ``TRUE''
coordinates of the beam through:
\begin{subequations}
\label{eq:oa2true}
\begin{eqnarray}
A_T &=& \Aoa - O_A - B_A/\sin(\Zoa)\\
Z_T &=& \Zoa - O_Z - B_Z + R,
\end{eqnarray}
\end{subequations}
where $R$ is a refraction angle discussed latter and $O_A$ and
$O_Z$ are azimuth and zenith distance offsets which are added if requested.

The condition for tracking a source at fixed offset defined
in azimuth--zenith-distance space is:
\begin{subequations}
\label{eq:track1}
\begin{eqnarray}
A_S + O_A &=& A_T \\
Z_S + O_Z &=& Z_T.
\end{eqnarray}
\end{subequations}
For the case when the receiver beam tracks the source we have: $O_A=0$
and $O_Z=0$.  In the version of the control system which has been used
to gather the pointing data presented in this work, the receiver beam
offset $B_A$ scales with the zenith distance as $B_A/\sin(\Zoa)$
(Eq.~\ref{eq:oa2true}). We will hereafter refer to this version as
{\fastTrack }.

While tracking, these coordinates are sensitive to switching from one
off-axis receiver to another, as the tracking telescope must
physically slightly reorient itself when a new offset receiver (and
therefore offset beam) is requested. At the same time the ``SET''
coordinates are, obviously, not sensitive to these changes.  The pair
of ``TRUE'' and ``SET'' coordinates is provided numerically to other
programs and to users via local network.  While tracking a source,
``TRUE'' and ``SET'' coordinates are made equal to within tracking
precision by a proportional--integral--derivative (PID) controller.

For each moment in time the position encoders read the actual
orientation of the telescope in its own coordinate system, that only
crudely approximates the horizontal coordinate system.  We will call
these coordinates $A_E$ and $Z_E$. In order to relate these coordinates to the
``TRUE'' (in-space) coordinates $(A_T, Z_T)$ when the telescope is tracking a
source, they need to be corrected for
(i) tilt of the vertical axis with respect to geodetic zenith,
(ii) skew of the elevation axis,
(iii) focus box offsets 
(iv) gravitational sag of the Cassegrain mirror
(v) rail deficiencies
(vi) and other effects resulting from construction defects that cannot be easily modeled.
Hereafter, all of these corrections are jointly called COR, and they
allow us to connect the coordinates of the telescope own coordinate system to the
horizontal, in-space coordinates of the selected beam
via the weather-dependent atmospheric refraction angle ${R>0}$:
\begin{widetext}
\begin{subequations}
\label{eq:conv}
\begin{eqnarray}
A_T + O_A + B_A/\sin(\Zoa) & = & A_E - C_A(A_E,Z_E) - T_A(A_E,Z_E) \\
Z_T + O_Z + B_Z           & = & Z_E - C_Z(A_E,Z_E) - T_Z(A_E,Z_E) + R(\Zoa,T,P,H),
\end{eqnarray}
\end{subequations}
\end{widetext}
where $R$ is a function of $\Zoa$ (zenith distance of the
optical axis in the horizontal coordinate system) and actual
temperature ($T$), pressure ($P$) and humidity ($H$), provided to the
control system from local meteorological station. The control system
also accommodates for additional position corrections $T_A$ and $T_Z$
which are taken from look-up tables, although these were not used
during the pointing campaign i.e. for the data considered in this work
$(T_A, T_Z)=(0,0)$.  Obviously, the pointing data discussed in
Sec.~\ref{sec:data}, have no user-defined offsets, hence in this
analysis $(O_A,O_Z)=(0,0)$.  The $C_A$ and $C_Z$ terms in
Eqs.~\ref{eq:conv} represent all of the COR effects. In this setting,
pointing imperfections result mainly from our ignorance about the true
orientation of the optical axis.  In the case when COR is not
exact, pointing measurements will indicate non-zero position
corrections at the source direction.  We can account for that by
rewriting Eqs.~\ref{eq:conv} as:
\begin{widetext}
\begin{subequations}
\label{eq:conv2}
\begin{eqnarray}
A_T + B_A/\sin(\Zoa) + \Delta_A &=& A_E - C_A(A_E,Z_E)  \\
Z_T + B_Z            + \Delta_Z &=& Z_E - C_Z(A_E,Z_E) + R(\Zoa,T,P,H),
\end{eqnarray}
\end{subequations}
\end{widetext}
where $(\Delta_A,\Delta_Z)$ are the measured position corrections.
Using Eqs.~\ref{eq:oa2true} this can be rewritten in terms of the coordinates of the telescope optical axis:
\begin{subequations}
\label{eq:conv3}
\begin{eqnarray}
\Aoa + \Delta_A &=& A_E - C_A(A_E,Z_E)  \\
\Zoa + \Delta_Z &=& Z_E - C_Z(A_E,Z_E).
\end{eqnarray}
\end{subequations}

Table~\ref{tab:coords} summarizes the coordinate systems and naming conventions
introduced in this section.
\begin{table*}[t]
\caption{List of coordinates and naming conventions.}
\begin{tabular}{p{3cm}p{1.5cm}p{13.5cm}}
\hline\hline

Control system name & Name & Comment\\
SET & $(\alpha_S,\delta_S)$ & Equatorial coordinates of the
selected source, calculated from user-provided coordinates for
a given epoch \\
SET & $(A_S,Z_S)$ & Actual (in-vacuum) horizontal coordinates of
the selected source, corresponding to $(\alpha_S,\delta_S)$ at the
actual UT1 time \\
TRUE & $(A_T,Z_T)$ & Actual (in-vacuum) horizontal coordinates of
the receiver beam\\
TRUE & $(\alpha_T,\delta_T)$ & Equatorial coordinates coordinates
of the receiver beam corresponding to $(A_T,Z_T)$\\
& $(\Aoa,\Zoa)$ & Horizontal coordinates of the telescope optical
axis.  These coordinates are calculated by the control system
from $(A_E,Z_E)$ with assumptions on pointing model (COR).
$\Aoa=A_E-C_A(A_E,Z_E)$ and $\Zoa=Z_E-C_Z(A_E,Z_E)$.
\\
& $(A_E,Z_E)$ & Coordinates of the optical axis in the
telescope own coordinate system that approximates horizontal coordinate
system with imperfections described by COR
(see. Sec.~\ref{sec:setting}).  These coordinates are read, but
not available to the user. \\
& $(C_A,C_Z)$ & pointing model corrections (COR) in $A_E$ azimuth and $Z_E$ zenith
distance\\
& $(T_A,T_Z)$ & Look-up table corrections in $A_E$ azimuth and
$Z_E$ zenith distance\\
& $R$ & Weather dependent radio refraction angle. $R>0$\\
& $(O_A,O_Z)$ & User-defined position offset, defined in the
horizontal coordinate system. $O_A$ increases westwards. $O_Z$
increases downwards.\\
& $(B_A,B_Z)$ & Receiver beam offset from optical axis defined
in the horizontal coordinate system.  $B_A$ depends on elevation
as $B_A=B_A(\Zoa=90^\circ)/\sin(\Zoa)$.  $B_A$ increases
westwards. $B_Z$ increases downwards.\\
& $(\Delta_A,\Delta_Z)$ & Position corrections measured using the
{\fastTrack } version of the control system that uses the
``Model 4c'' pointing corrections model (see Sec.~\ref{sec:model}).
These position corrections need to be applied at the top of the
used pointing model (COR) to point a beam at the requested source.\\
\hline\hline
\end{tabular}
\label{tab:coords}
\end{table*}

\section{Pointing model}
\label{sec:model}
In the {\fastTrack } version of the control system the position
corrections (COR) are defined by an analytic pointing model called
``Model 4c'' \citep{Borkowski2004,Borkowski2006} which is based on the
model derived by \cite{Himwich1993}. This model was used during the
pointing observations (Sec.~\ref{sec:observations}).

The problem of improving the {\RT} pointing precision is
a matter of finding new analytic models
as a replacement for $C_A(A_E,Z_E)$ and $C_Z(A_E,Z_E)$ functions
(Eqs.~\ref{eq:conv3}) such that they minimize the position
corrections $(\Delta_A,\Delta_Z)$ amended by the pointing model
used to measure them. For this purpose we define:
\begin{subequations}
\begin{eqnarray}
\epsilon_A(A_E,Z_E) &=& \Delta A+ C_A(A_E,Z_E) \\
\epsilon_Z(A_E,Z_E) &=& \Delta Z+ C_Z(A_E,Z_E).
\end{eqnarray}
\label{eq:epsilon} 
\end{subequations}

We also define a new pointing corrections model for azimuth $M_A({\bf
p},A_E,Z_E)$ and zenith distance $M_Z({\bf q},A_E,Z_E)$ and
we fit pointing measurements by minimizing:
\begin{subequations}
\begin{eqnarray}
\chi^2_A &=& \sum\limits_{i=1}\limits^N \frac{w_i^2}{\sigma^2_{A_i}} \Big(\epsilon_{A_i} - \ModelA{Y}({\bf p},A_E,Z_E)\Big)^2\\
\chi^2_Z &=& \sum\limits_{i=1}\limits^N \frac{w_i^2}{\sigma^2_{Z_i}} \Big(\epsilon_{Z_i} - \ModelZ{Y}({\bf q},A_E,Z_E)\Big)^2,
\end{eqnarray}
\label{eq:chisq} 
\end{subequations}
where ${\bf p}$ and ${\bf q}$ are parameters of the $\ModelA{Y}$ and
$\ModelZ{Y}$ models respectively, $(Y)$ indicates the version of the
model used for fitting, $N$ is the number of pointing measurements,
$\sigma^2_{Z_i}$ quantifies the $i$'th measurement noise level
and $w_i\sim S_i$ is proportional to the radio source
flux density ($S_i$). In practice, we assume $w_i/\sigma_i =
\log({\rm SNR_i})$, where SNR is the signal to noise ratio
estimated from each pointing measurement.

The choice of the weighting function is to give stronger $\chi^2$
contributions from more reliable measurements while accounting for the
steep spectrum of the distribution of the intrinsic flux densities in
the observed radio source population. This choice may impact the
reconstructed confidence intervals, but as long as the numerical
precision is not a concern (and for the assumed Markov chain convergence
criteria) the choice should not affect the best fit solutions. In this
work, we are only concerned with finding the best fitting model through
$\chi^2$ minimization, and we do not reconstruct the parameter posterior
distributions.

In the {\fastTrack } version of the control system, the coordinates $(A_E, Z_E)$
and the values $C_A(A_E,Z_E)$ and $C_Z(A_E,Z_E)$ for any given
measurement are not directly available from observations.
We therefore use the following approximation:
\begin{subequations}
\begin{eqnarray}
\epsilon_A(A_E,Z_E)  &\approx& \Delta A+ C_A(\Aoa,\Zoa) \\
\epsilon_Z(A_E,Z_E)  &\approx& \Delta Z+ C_Z(\Aoa,\Zoa).
\end{eqnarray}
\label{eq:approx} 
\end{subequations}

In order to derive the $\chi^2$ values for any given set of parameter
values, a similar approximation is used to
calculate the new pointing model corrections ($M_A$ and $M_Z$) at
$(A_E,Z_E)$.
While the transformation from
$(A_T, Z_T)$ to $(\Aoa, \Zoa)$ coordinates is possible by
reconstructing the weather dependent refraction history
(Eqs.~\ref{eq:oa2true}), in the current work we perform this
transformation using the mean refraction model
(Eqs.~\ref{eq:T2OA}) which is sufficient given the quality of the
present data, and the fact that the data cover $Z_T< 80^\circ$
where $R< 0.15^\circ$. The COR corrections in this range are
even smaller.

When calculating $\chi^2$ values we make at least two implicit
approximations that should be addressed. One results from calculating
the model position correction values at wrong directions: $(\Aoa, \Zoa)$ rather
than $(A_E, Z_E)$, and the other stems from associating the measured
corrections $(\Delta_A,\Delta_Z)$, to wrong directions: 
$(\Aoa,\Zoa)$, rather than $(A_E, Z_E)$.  This is expressed in
Eqs.~\ref{eq:approx}.  The approximation is justified by the fact that
the fastest recorded changes of the position corrections are about
$0.003$ deg/deg, which when converted to the angular scales of the
differences between $(A_E, Z_E)$ and $(A_T, Z_T)$ coordinates (even at
those lowest elevations), the approximation gives errors of the order
$0.0003^\circ$ at the most.  In practice, the errors are much smaller
because ``Model 4c'' does not model the fastest variations of the
corrections associated with rail irregularities, and the rate of
correction changes in this model is actually much smaller than the
measured value.  Therefore, for the current data the error due to this
approximation is not important.

Except for the rail irregularity, the effects associated with
pointing corrections described in Sec.~\ref{sec:setting} can be
accurately modeled by slightly modified formulas derived in
\cite{Borkowski2004}:
\begin{widetext}
\begin{subequations}
\begin{eqnarray}
\ModelA{4e}({\bf p},\Aoa,\Zoa) &=& A_0 + \big(\xi_A \sin(\Aoa) - \zeta_A \cos(\Aoa) + \sigma\big) \cot(\Zoa) + \frac{\beta}{\sin(\Zoa)}+ p_1 \sin(2 \Aoa) \\
&+& p_{2}\cos(2 \Aoa) + p_3\sin(3\Aoa)\cos(\Zoa) + p_4\cos(\Aoa/4)\sin(\Zoa)\nonumber\\
\ModelZ{4e}({\bf q},\Aoa,\Zoa) &=& Z_0 + \xi_Z \cos(\Aoa) + \zeta_Z\sin(\Aoa) + \gamma\sin(\Zoa) + q_1 \cos \Zoa + q_{2}\sin (2 \Aoa) + q_{3}\cos (2 \Aoa)
\end{eqnarray}
\label{eq:model4e} 
\end{subequations}
\end{widetext}
where
$\protect{{\bf p}=\{A_0,\xi_A,\zeta_A,\sigma,\beta,p_1, p_{2},p_3,p_4\}}$ and
$\protect{{\bf q}=\{Z_0,\xi_Z,\zeta_Z,\gamma,q_1, q_2, q_3\}}$
are the model parameters.
This 16-parameter model is a small angle limit of the exact
``Model 4c'', but extended by an additional ad-hoc parameter $q_3$,
which proves to be useful in mitigating large-scale trends in
zenith distance residuals.  The model has a simpler formulation
than the full ``Model 4c'' ($\Model{4c}$), but retains a very high
compatibility with it
(Borkowski, 2016, private communication).
We will hereafter refer to this model as ``Model 4e'' or $\Model{4e}$.
In the original version of model $\Model{4c}$, the parameter $q_3=0$
and the equations are coupled since $\xi_A=\xi_Z=\xi$ and
$\zeta_A=\zeta_Z=\zeta$, which is physically justified.

The $A_0$ and $Z_0$ are simple constant offsets, $\gamma$ calibrates
the gravitational sag of the secondary mirror and its supports, $\xi$
and $\zeta$ parameters define tilt angles of the telescope azimuth axis
towards local meridian and towards the West (the so-called tilt-over
and tilt-out components), $\sigma$ defines angle between the elevation
axis and the plane of local horizon, and $\beta$ is the angle between
plane perpendicular to elevation axis and the optical axis. Seven
ad-hoc terms ($p_1$, $p_2$, $p_3$, $p_4$, and $q_1$, $q_2$, $q_3$) are inserted empirically to accommodate for corrections that
cannot be modeled by the aforementioned construction imperfections, but
may be related with e.g. eccentricity of gearwheels etc.

When the equations are coupled ($\xi_A=\xi_Z=\xi$ and
$\zeta_A=\zeta_Z=\zeta$) the two models ($M_A$ and $M_Z$) can be
fitted jointly by minimizing:
\begin{equation}
\chi^2 = \chi^2_A+\chi^2_Z
\label{eq:chisqSum} 
\end{equation}
however, we find that the resulting best-fit model yields residuals a
factor of ${\sim}1.7$ times larger in zenith distance and ${\sim}1.1$
times larger in azimuth, than in the case when the two parameters are
allowed to differ between the models for each of the two coordinates.
We will therefore use the model with 16 parameters as defined in
Eqs.~\ref{eq:model4e}, even though it is not as well physically
motivated.

\section{Observations}
\label{sec:observations}
Between April and October 2016 we carried out a pointing campaign
using OCRA-p \citep{Browne2000}, a 30-GHz dual beam, beam-switched
receiver with half power beamwidth $\theta_{\rm OCRA}\approx
1.2'$ and with the reference beam offset
\begin{equation}
(B_A,B_Z)_{\rm OCRAp}=(0.432^\circ,-0.0720^\circ).
\label{eq:roh}
\end{equation}
During the observations the pointing model ``Model 4c'' (Sec.~\ref{sec:model}) was used along
with the {\fastTrack} version of the control system. The
observations were carried out using the ``ocraToolkit'' 
software package, designed for OCRA-SZ observational project
\citep{Lancaster2011,Lew2015,Lew2016b}.

\begin{figure}
\centering
\includegraphics[width=0.5\textwidth]{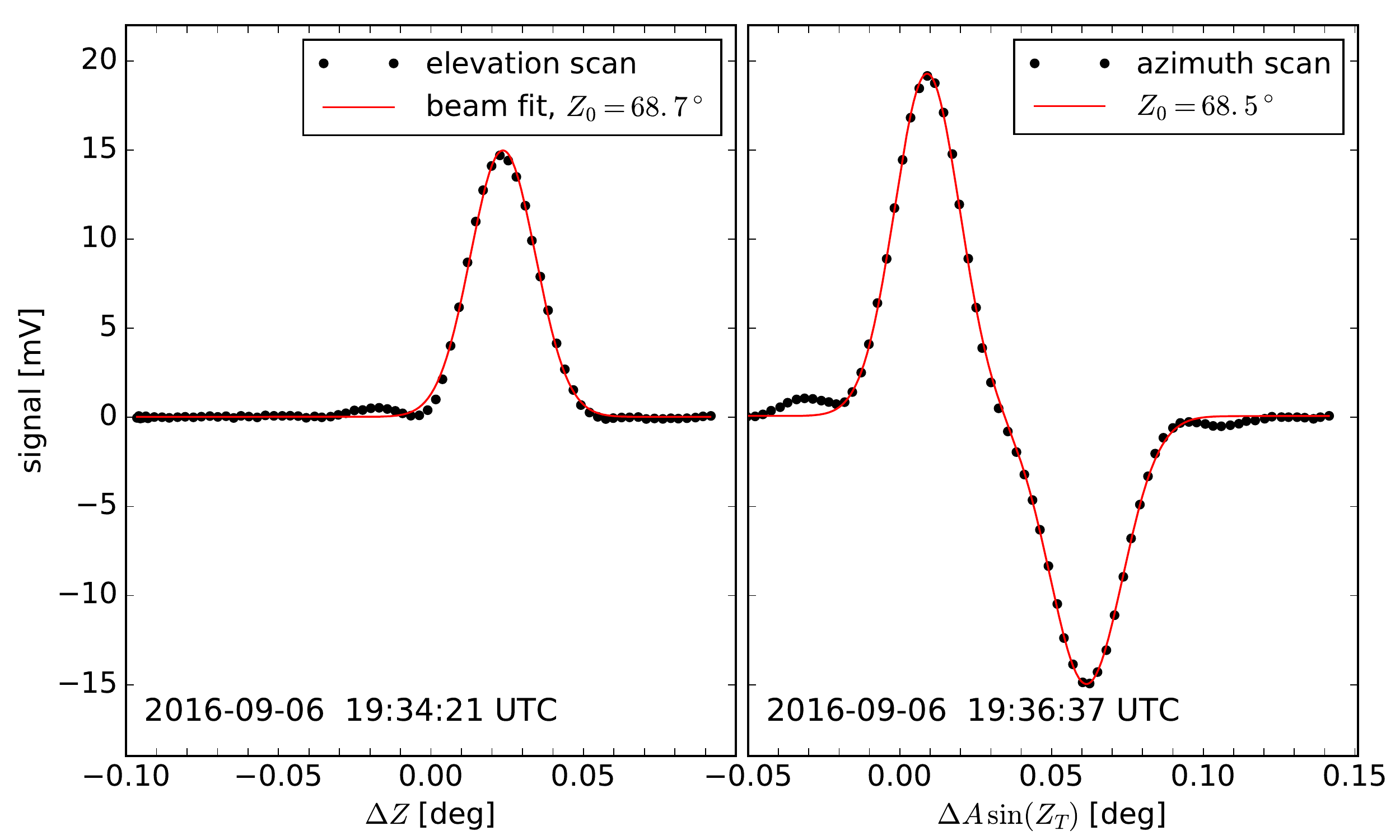}
\caption{A high signal to noise (${\rm SNR}\approx 510$) cross-scan observation of 3C 84
with a linear drift removed. First, elevation scan is performed (left)
and it is followed by azimuth scan (right), accounting for the position
correction from the elevation scan. Side lobes of up to 5\%
are asymmetric due to receiver beam offset (Eq.~\ref{eq:roh}).}
\label{fig:cross_scan}
\end{figure}

\begin{figure*}
\centering
\includegraphics[width=0.85\textwidth]{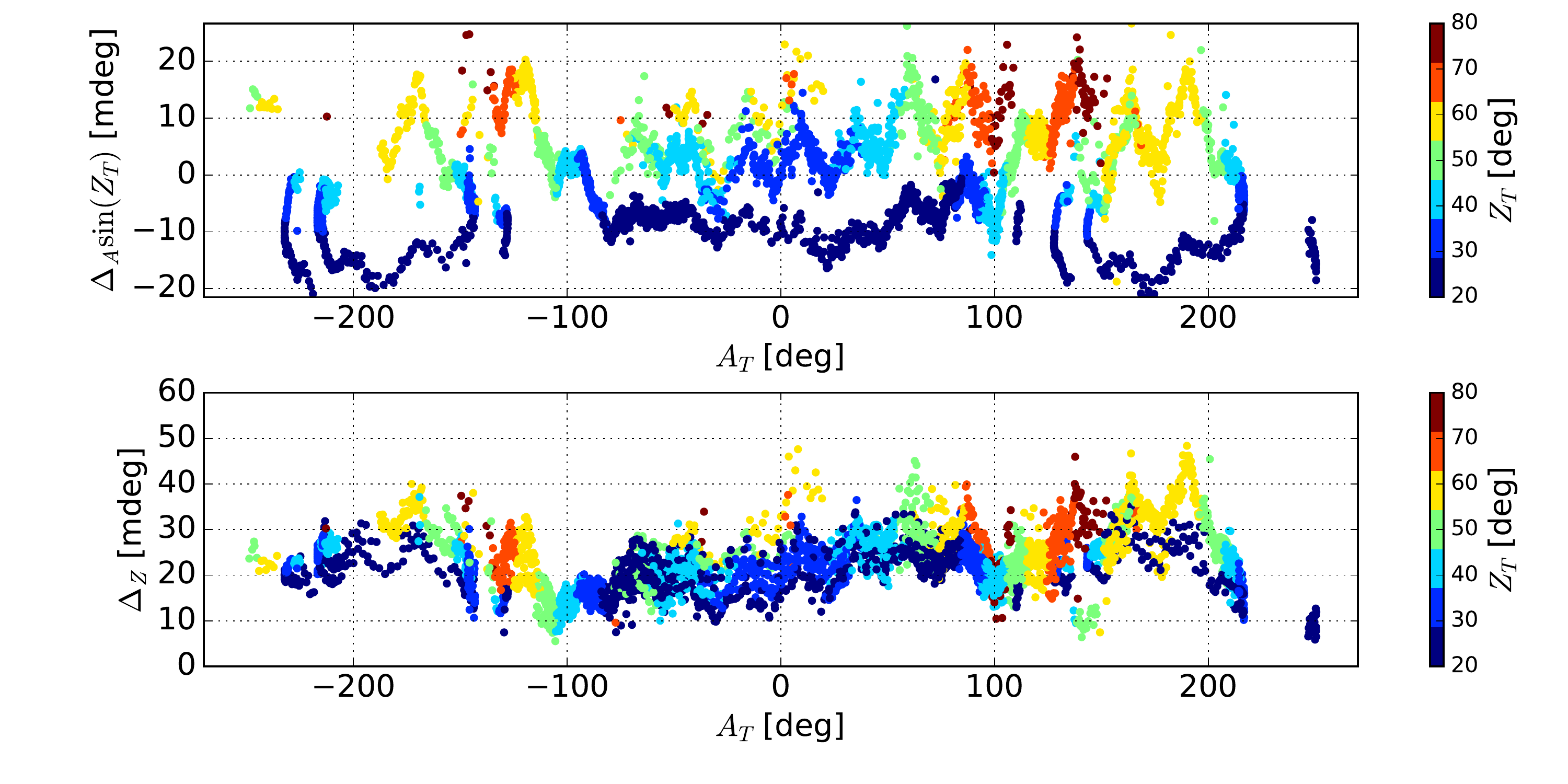}
\caption{Pointing corrections in azimuth ({\it top}) and zenith
distance ({\it bottom}) obtained from pointing observations as a function
of azimuth and zenith distance.}
\label{fig:corrections}
\end{figure*}

For any given radio source, we perform a cross-scan
(Fig.~\ref{fig:cross_scan}) as previously discussed in \cite{Lowe2007}
and \cite{Gawronski2010}, and we extract the position corrections
$(\Delta_A,\Delta_Z)$ (Fig.~\ref{fig:corrections}) by fitting gaussian
and double gaussian functions to the data points after having removed
drifts arising due to atmospheric effects \citep{Lew2016}.  The
averaged cross-scan data have 1-s time resolution.

We observed a sample of 26 distinct radio sources.  The sample is
composed mostly of active galactic nuclei (AGNs) including:
intermediate to high redshift ($0.2<z<2.5$) quasars (14), and low to
intermediate redshift ($0.017<z<0.72$) Seyfert galaxies (6), BL Lac
objects (4), a radio galaxy (1) and the NGC 7027 planetary nebula, all
of which should appear a point-like with OCRA-p beam. However, the
majority of observations (92\%) were performed using a small sub-group
of brightest sources (9).  By analyzing the the quality of the fits of
the model beam to the data, as well as the repeatability of the
corrections within ranges where they vary slowly, we estimate that for
the data with high SNR the uncertainty of position correction
determination is small $O(10^{-4})$ deg, but possibly up to
${\approx}0.004$ deg for poor weather or low SNR.

\begin{figure}
\centering
\includegraphics[width=0.5\textwidth]{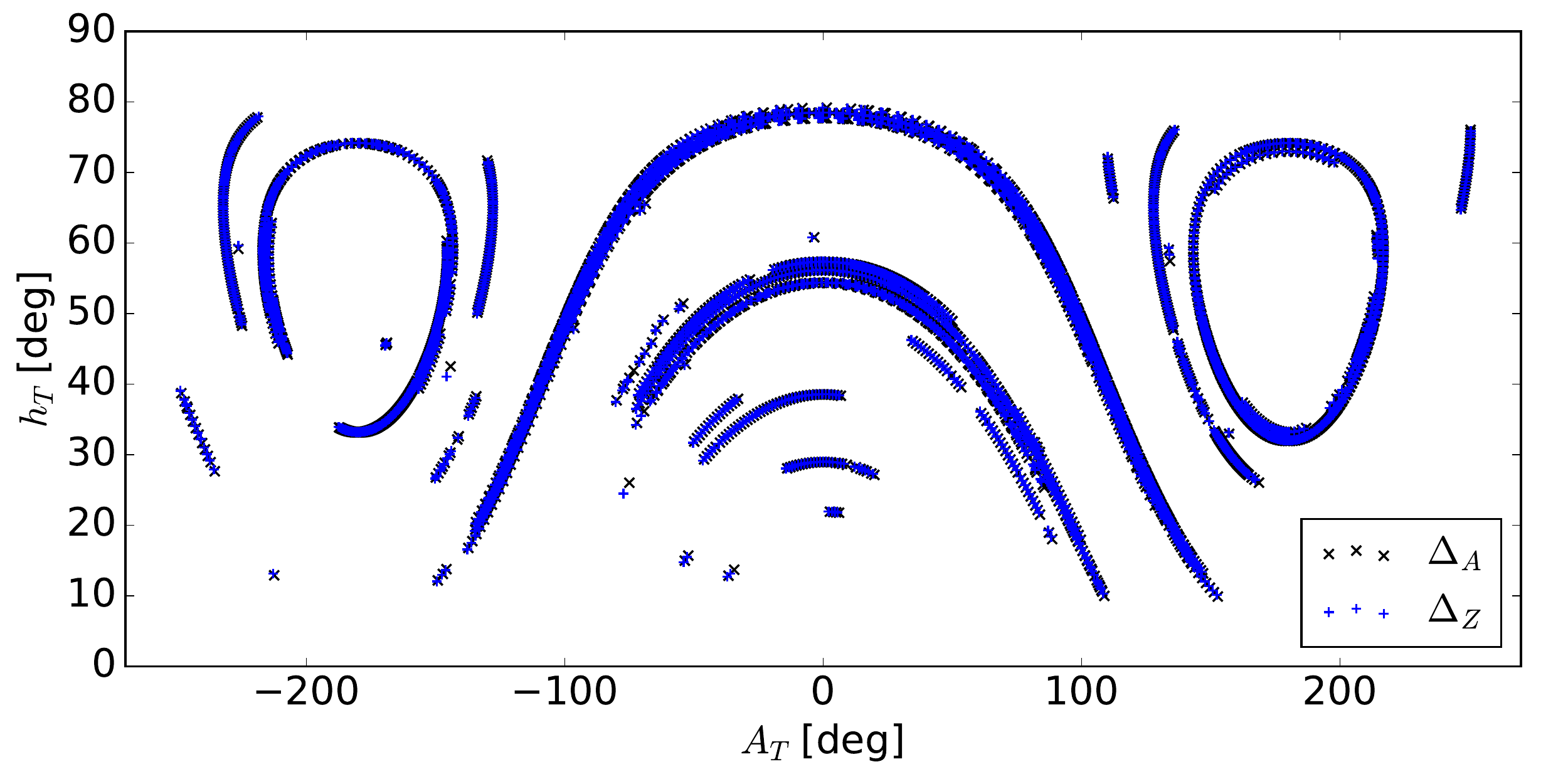}
\caption{Distribution of \stageTwo pointing measurements in
azimuth-elevation plane (see Tab.~\ref{tab:coords}).  In most cases
the direction of azimuth correction of a given cross-scan is close
to the direction of zenith distance correction of the same
cross-scan. In some records, this is not the case due to
e.g. failure in finding the best-fit position correction in the
first iteration. In such cases, the scan is repeated along that
coordinate until a satisfactory fitting is achieved a few minutes
later.}
\label{fig:sky_coverage}
\end{figure}

\begin{figure}
\centering
\includegraphics[width=0.4\textwidth]{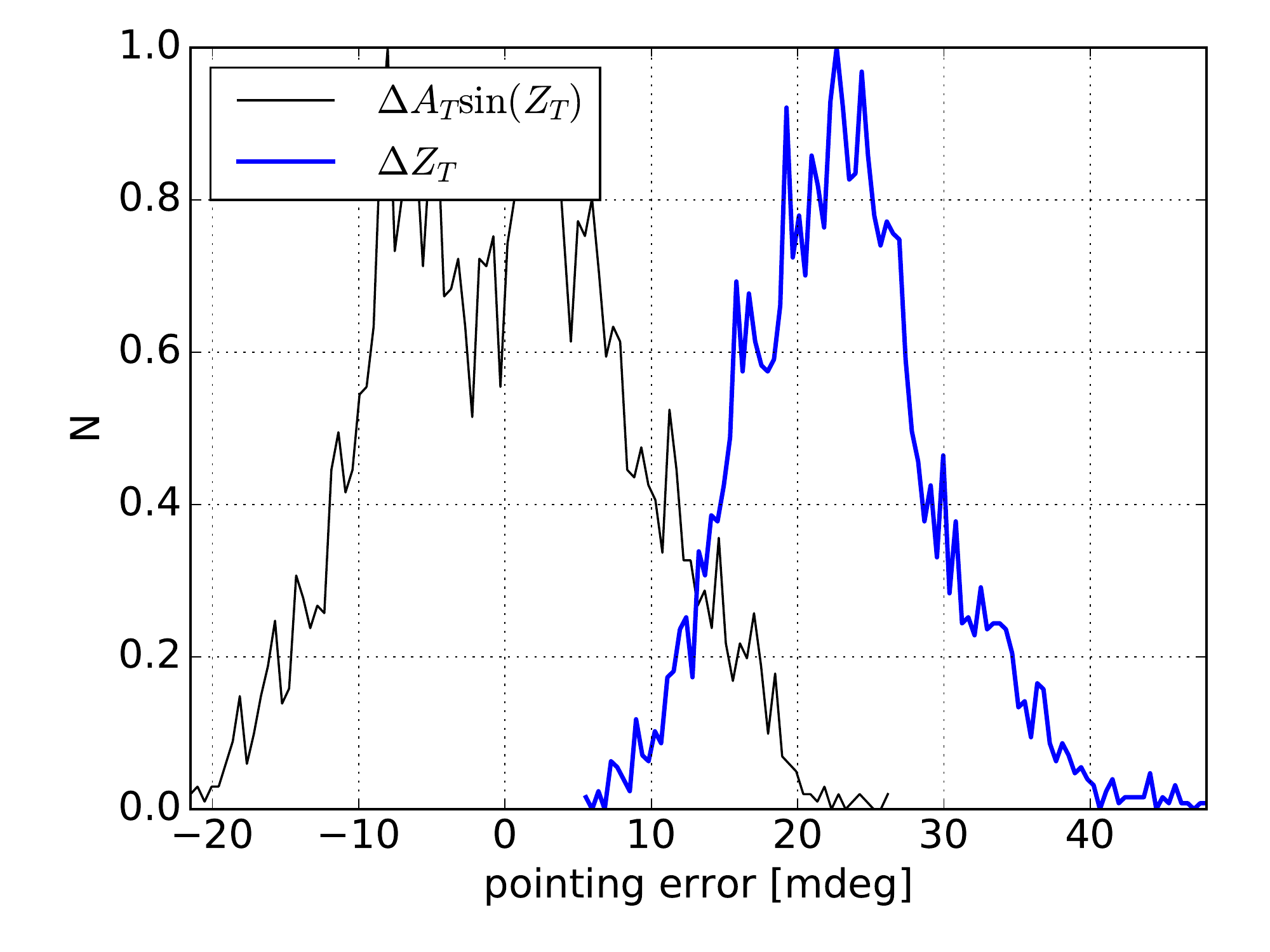}
\caption{Distribution of pointing errors from
Fig.~\ref{fig:corrections} (Sec.~\ref{sec:data}). The root mean
square (RMS) values of the pointing corrections are given in
Table~\ref{tab:pointing}.}
\label{fig:fastTrackPointingError}
\end{figure}

For the pointing campaign the control system has been modified to
account for a number of astrometric effects which are relevant at the
milli-degree pointing accuracy. The modifications include: (i)
introducing UT1 time scale\footnote{All time-related operations in the
control system are based on UT1 time scale. The UT1 time is updated
regularly using the data available online at
\url{http://maia.usno.navy.mil/ser7/mark3.out}. } which gives an
effect of ${<}0.6$s in hour angle, (ii) adapting the nutation model as
implemented in the NOVAS library
\citep{Kaplan2012}\footnote{\url{http://aa.usno.navy.mil/software/novas/novas_info.php}. While
the previous implementation of the nutation was consistent with
NOVAS library at the level ${<}10^{-5}$ deg, it was not used due to
bugs introduced during porting the code from FORTRAN to C.} which
affects pointing at the level ${\lesssim}5$ mdeg, (iii) introducing
weather dependent radio refraction as implemented in SLA library
\citep{Wallace1994} with weather data based on real-time
readouts from the local meteorological station (effect of ${\lesssim}50$ mdeg
for $Z<80$ and relevant meteorological conditions), (iv) enabling
annual aberration in coordinate transformations\footnote{The diurnal
aberration and polar motion are currently neglected.} (effect of
${\lesssim}6$ mdeg), and (v) correcting the telescope geodetic
coordinates to match those obtained from a geodetic VLBI experiment
\citep{Charlot2001}, which gives an effect of ${\approx}0.72$ s in
hour angle.

We have also improved the observing strategy that exploits, the
possibility of controlling the velocity of the telescope drives in
azimuth and elevation in order to scan a source along one coordinate
while tracking, rather than imposing a series of fixed offsets that
are reached along an unpredictable trajectory. This also allows us to
directly control the cross-scan speed.  It was estimated that for any
given cross-scan these improvements provide about 10\% larger receiver
response due to the radio source passing closer to the beam center as
the telescope sweeps the sky during the azimuth scan.

Another improvement over the previous pointing observations, comes
from a higher time resolution (roughly 8 Hz) of the telescope position
readouts obtained from the RTLinux based control system (a factor of
${\sim}8$ improvement) and also from the way the position data are
matched to the signal stream -- i.e. by using an interpolation rather
than the nearest neighbor approach.

\section{Data processing}
\label{sec:data}

\begin{figure*}
\centering
\includegraphics[width=0.49\textwidth]{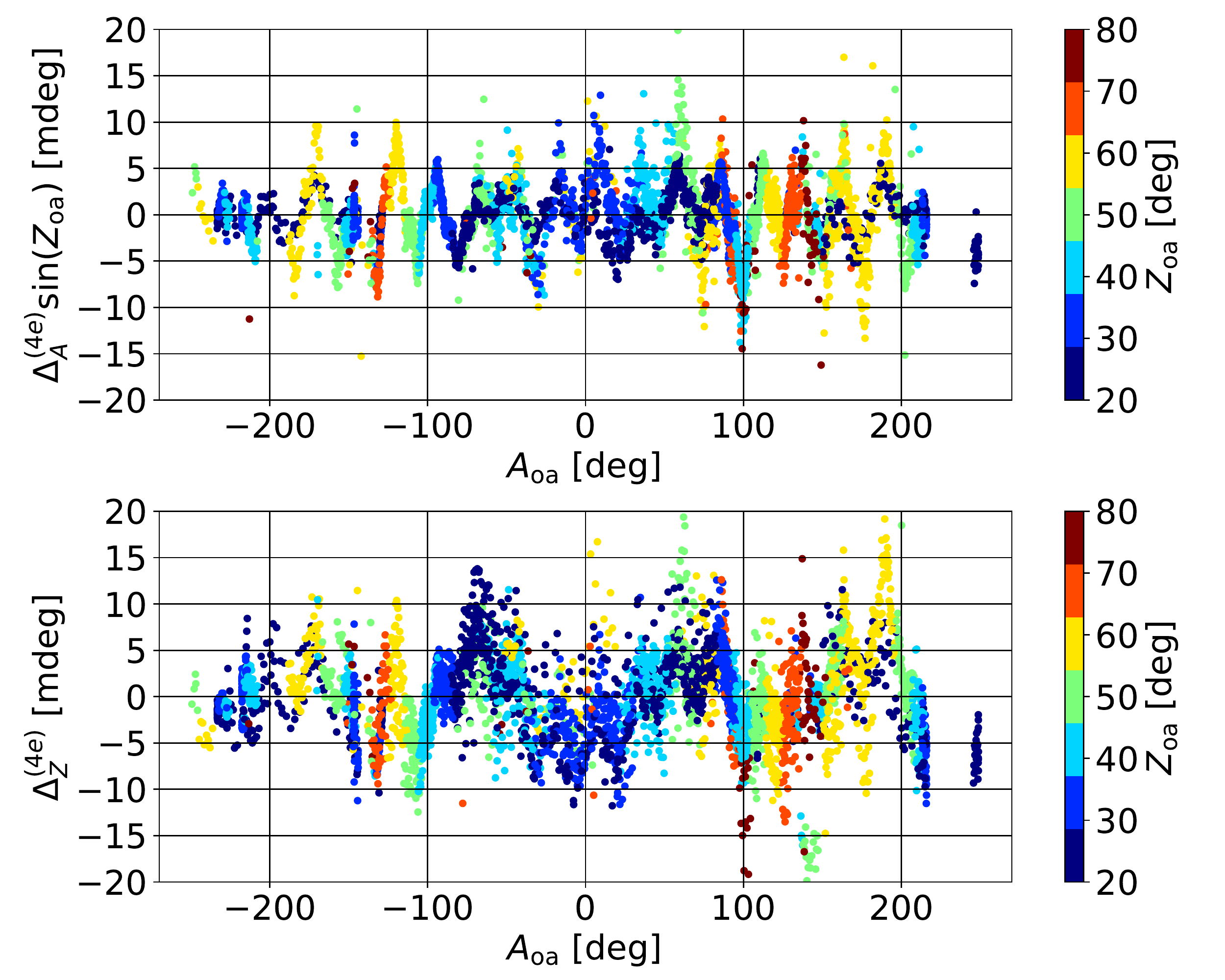}
\includegraphics[width=0.49\textwidth]{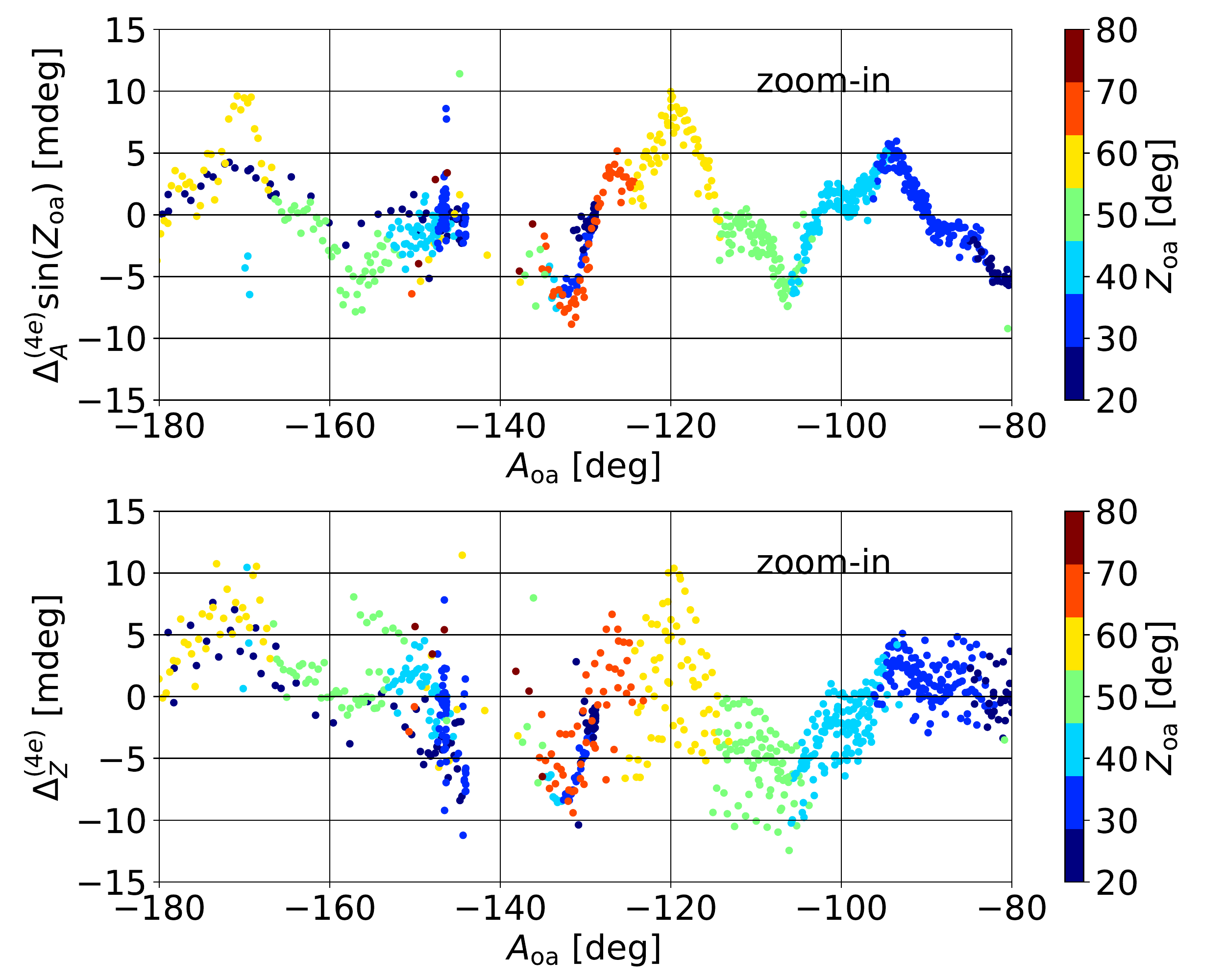}
\caption{Residuals between the pointing data and the best fit model
``Model 4e'' {\it (left)} and a zoom-in region {\it (right)}. Notice,
that for certain azimuth ranges the position corrections differ
depending on zenith distance of the source used for measurement,
whereas in other ranges there is a good overlap regardless of the
source elevation.}
\label{fig:model4eresid}
\end{figure*}

\begin{figure}
\centering
\includegraphics[width=0.4\textwidth]{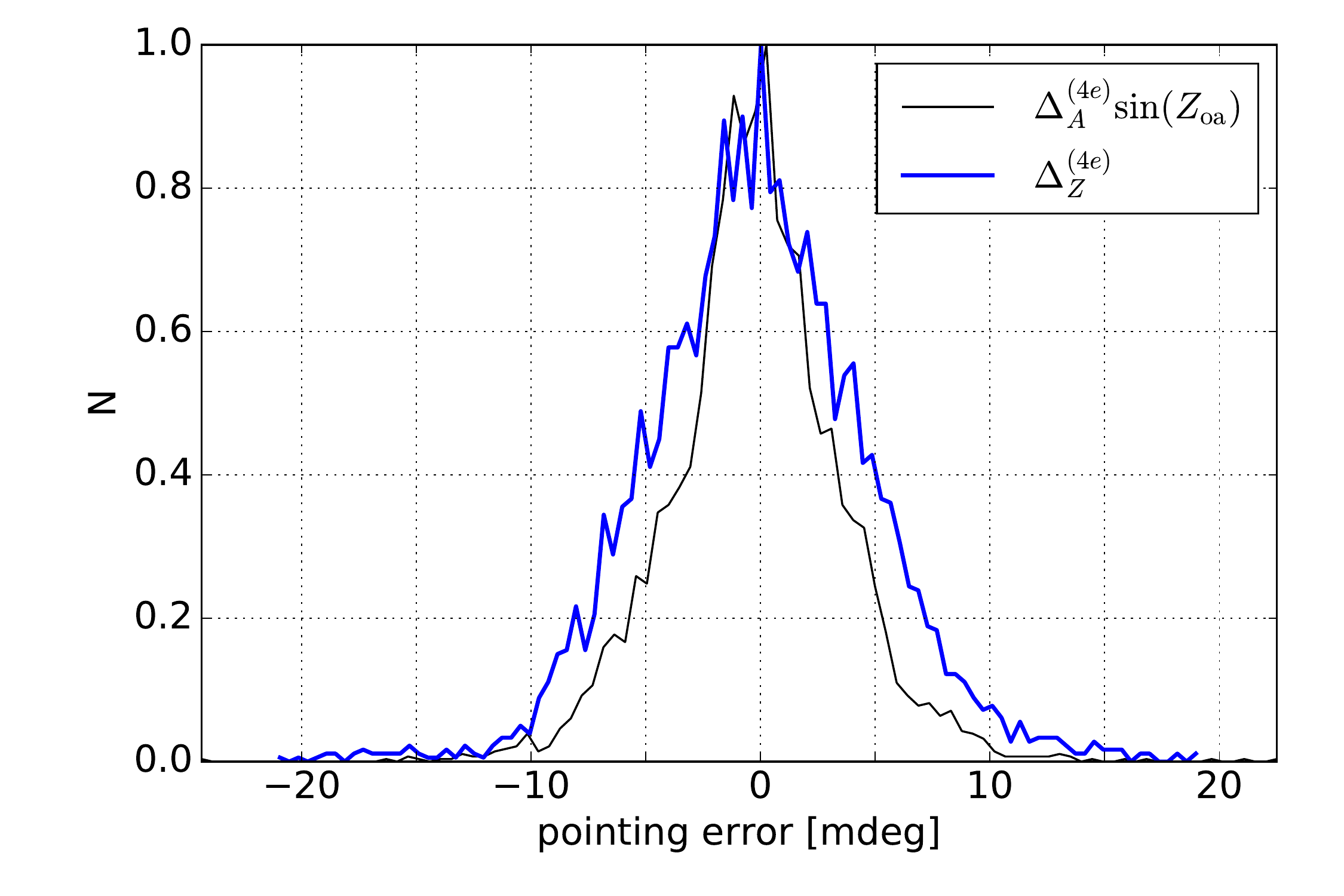}
\caption{Distribution of ``Model 4e'' residuals from
Fig.~\ref{fig:model4eresid}. The RMS values of the residuals are given
in Table~\ref{tab:pointing}.}
\label{fig:model4eresidDistr}
\end{figure}

The pointing data are a compilation of cross-scan observations of
selected radio sources, observed over ranges of hour angles with a
typical time lapse between the adjacent observations of ${\sim}5$
minutes.  Depending on the type of observational program, the time
elapsed between subsequent cross-scan observations of the same source
can increase to several minutes.  Each scan record contains
$\{A_T,Z_T,\alpha_T,\delta_T,t,\Delta, {\rm SNR}\}$
(Table~\ref{tab:coords}), where $t$ is the UT1 time of the source
maximal response, $\Delta$ is the derived position offset in azimuth
or zenith distance and SNR is a signal-to-noise estimator that we use
to weight contributions to the $\chi^2$ 
(Eqs.~\ref{eq:chisq}).  Each cross-scan observation contains two
such records: one for the azimuth scan in which case
$\Delta\equiv\Delta_A$ and one for the elevation scan in which case
$\Delta\equiv\Delta_Z$.

The cross-scan data are pre-processed before the analysis in several
stages.  On 14 June 2016 a number of important astrometric
corrections have been implemented into the control system
(Sec.~\ref{sec:observations}), therefore in \stageOne we select only
the data gathered after that date.

In \stageTwo we further screen this data to remove obvious
outliers (Fig.~\ref{fig:corrections}).  This is done by requiring
\begin{subequations}
\begin{eqnarray}
&|\Delta_A|& < 0.03^\circ\\
0.005^\circ < &\Delta_Z& < 0.05^\circ 
\end{eqnarray}
\label{eq:stage2sel}
\end{subequations}
This condition leaves 4076 pointing measurements that fully cover the
entire range of azimuths and the observationally useful range of zenith distances
(Fig.~\ref{fig:sky_coverage}). Clearly, for any given $A_T$ the pointing
corrections depend on $Z_T$ as well.  However, they do so consistently: i.e.
for a given $A_T$, negative $\Delta_A$ corrections reside
typically at low zenith distances (and vice-versa), but not at low and
high zenith distances simultaneously, which would hint on data inconsistency
or time-dependent effects. The same seems to be true for
$\Delta_Z$ corrections.  A statistic of pointing precision of the
{\fastTrack } version of the control system is shown in
Fig.~\ref{fig:fastTrackPointingError}.

In the \stageThree of data pre-processing coordinates of each
pointing measurement are transformed from $(A_T, Z_T)$ to $(\Aoa,
\Zoa)$ using OCRA-p receiver beam offset (Eq.~\ref{eq:roh}) and
the mean optical atmospheric refraction according to:
\begin{subequations}
\begin{eqnarray}
\Aoa_i &=& A_{Ti} + B_A^{\rm OCRAp}/\sin(\Zoa_i)\\
\Zoa_i &=& Z_{Ti} + B_Z^{\rm OCRAp} - R(Z_{Ti}).
\end{eqnarray}
\label{eq:T2OA}
\end{subequations}
Using the approximation
given in Eqs.~\ref{eq:approx} we calculate
$\epsilon_{A_i}(\Aoa,\Zoa)$ and $\epsilon_{Z_i}(\Aoa,\Zoa)$.
In Eq.~\ref{eq:T2OA}b we use formula for optical refraction even though radio and weather-dependent refraction
model was used during the observations. This shortcoming has little effect on reconstructing $\Zoa$ directions
as explained in Sec.~\ref{sec:model}.

\section{Finding the best-fit model}
\label{sec:bestfit4e}

We fit the 16-parameter model as defined in Eqs.~\ref{eq:model4e}
to the pointing data discussed in Sec.~\ref{sec:data} using
Monte-Carlo Markov-Chain (MCMC) approach, combined with simulated
annealing (SA) algorithm.
A typical MCMC chain with a moderate cooling rate takes about
20\,000 to  40\,000 steps before a converged solution is found.
We assume a flat initial prior distribution for each parameter.

The residuals between the pointing data and the best-fit ``Model 4e'':
\begin{subequations}
\begin{eqnarray}
\Delta^{(4e)}_{A}&=&\epsilon_A - \ModelA{4e}({\bf p},\Aoa,\Zoa)\\
\Delta^{(4e)}_{Z}&=& \epsilon_Z - \ModelZ{4e}({\bf q},\Aoa,\Zoa)
\end{eqnarray}
\label{eq:resid} 
\end{subequations}
are shown in Fig.~\ref{fig:model4eresid}.

The figure clearly shows the possibility of improving the telescope
pointing precision. The zoom-in panels also show, that what looks like
a noise (left plots in Fig.~\ref{fig:model4eresid}) actually has a
fine structure that is resolved with the current quality of the data
(e.g. Fig.~\ref{fig:model4eresid} top-right).  The statistical error
of a pointing measurement, $O(10^{-4})$ degree, is much smaller than
the systematic errors still present in $\Delta^{(4e)}$
residuals and clearly, model $\ModelA{4e}$ is unable to fit them.  This
indicates that the pointing model (Eqs.~\ref{eq:model4e}) can be
further improved. The residuals also confirm that the rate of variation
of the azimuth corrections is ${<}3\, {\rm mdeg/deg}$.

It is clear that for any given range of azimuths and/or elevations the
dispersion of the zenith distance pointing corrections is larger, than
it is in the case of azimuth corrections (Fig.~\ref{fig:model4eresid}).
This may indicate problems
of repeatability of the measurements (e.g. resulting from the
stability of the suspension of the secondary mirror). For certain
ranges of azimuths, the azimuth or elevation position corrections may
vary by as much as ${\sim}0.015^\circ$ for two distinct elevations
or distinct measurement dates. This also hints that some effects
yet unaccounted for may play some role.

The best fit parameter values in scientific notation in degrees yield:
\begin{subequations}
\begin{eqnarray}
{\bf p}&=&\{ 3.414471{\rm e}{-03}, -9.148816{\rm e}{-04}, \\
&& -2.298499{\rm e}{-03},  1.886013{\rm e}{-02}, \nonumber\\
&& -4.381365{\rm e}{-02},  -1.035695{\rm e}{-02}, \nonumber\\
&& 8.263103{\rm e}{-03},  -7.497834{\rm e}{-03}, \nonumber\\
&& 5.051955{\rm e}{-03} \}\nonumber\\
{\bf q}&=&\{ 8.268194{\rm e}{-02}, 4.271137{\rm e}{-05}, \\
&& 2.704925{\rm e}{-04},  -8.758806{\rm e}{-04}, \nonumber\\
&& -3.489975{\rm e}{-02},  -4.142412{\rm e}{-03}, \nonumber\\
&& 3.697197{\rm e}{-03} \}\nonumber
\end{eqnarray}
\end{subequations}

The anticipated improvement due to introducing model $\Model{4e}$ is shown
in Fig.~\ref{fig:model4eresidDistr} and in Table~\ref{tab:pointing}.
\section{Modeling rail surface irregularities}
\label{sec:model5}
\begin{figure*}
\centering
\includegraphics[width=0.5\textwidth]{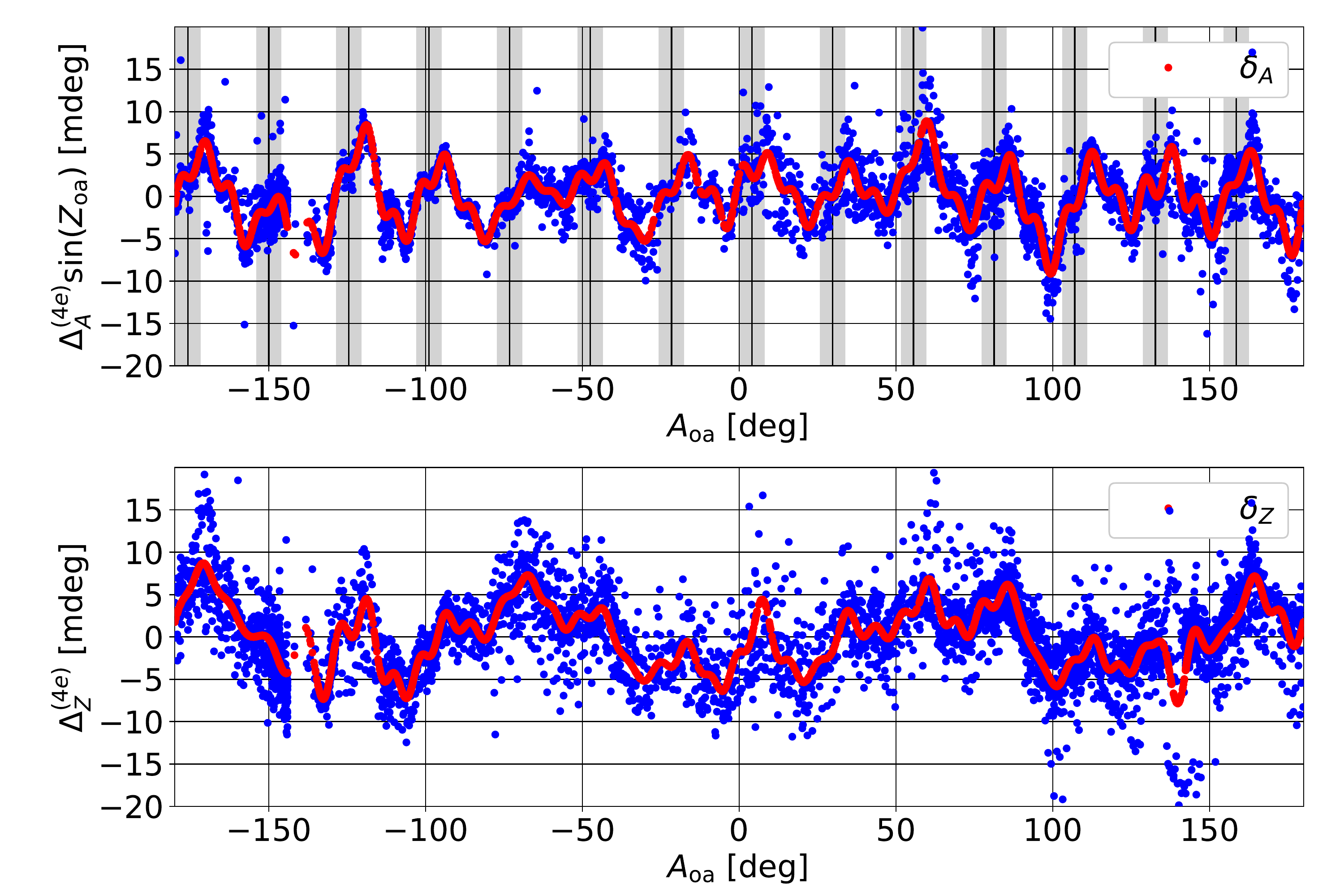}
\includegraphics[width=0.49\textwidth]{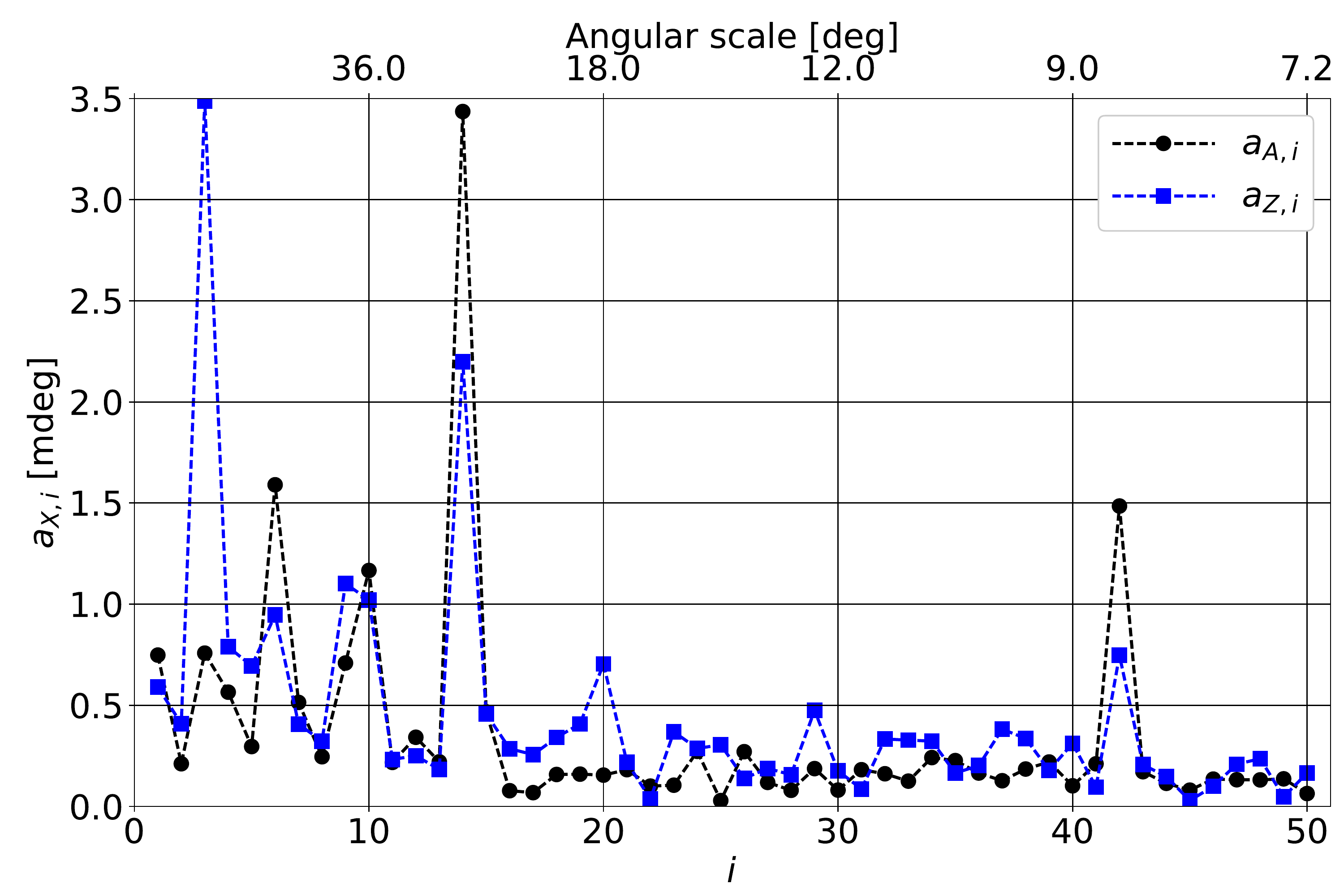}
\caption{{\em (Left)} Residuals between the pointing data and the best fit
model ``Model 4e'' wrapped into $A=[-180^\circ,180^\circ)$ range
(blue) and a Fourier expansion fit based on Eqs.~\ref{eq:model5fit}
(red). Vertical lines in the top panel mark the azimuths of the
rim welding points, and the shaded bands (${\approx}7.6^\circ$ wide)
correspond to the angular separation of the wheels in each of the four
trolleys supporting the telescope.
{\em (Right)} Fourier expansion coefficients $a_{X,i}$ as
defined in Eq.~\ref{eq:model5fit}. High amplitude modes clearly reach
into small angular scales, and many of the weaker modes
have their phases strongly correlated, which builds up the integrated corrections. } 
\label{fig:model4efit}
\end{figure*}

The fluctuations of $\Delta^{(4e)}$ are
largely independent from the zenith distance coordinate (Fig.~\ref{fig:model4eresid}),
thus introducing
further improvements simpler, as compared to the case when the
corrections or residuals depend on azimuth and elevation at the same time.

In order to further improve ``Model 4e'', we
(i) wrap $\Big(\Delta^{(4e)}_{A}\sin(\Zoa), \Delta^{(4e)}_{Z}\Big)$
residuals into $A=[-180^\circ, 180^\circ)$ range,
(ii) densely resample the corrections on a uniform grid using a linear
interpolation, and
(iii) decompose them into Fourier series.
The Fourier expansion yields:
\begin{equation}
\delta_X(\Aoa) = \delta_{X0} + \sum\limits_{i=1}^{N_F} a_{X,i} \sin\Big(\frac{2\pi}{T_{X,i}}(180-\Aoa) + \phi_{X,i}\Big)
\label{eq:model5fit} 
\end{equation}
where $X=A$ for azimuth residuals and $X=Z$ for zenith distance
residuals, $\delta_{X0}=\langle\Delta_{X_i}s_{X_i}\rangle_i \approx 0$
where $s_{X_i}=\sin{\Zoa_{,i}}$ for $X=A$ and $s_{X_i}=1$ for $X=Z$
and $\Delta_{X_i}$ is the $i$'th residual calculated using
Eqs.~\ref{eq:resid}.  We find that $N_F=50$ gives a reasonable fit to
the model $\Model{4e}$ residuals (Fig.~\ref{fig:model4efit}) and allows
us to reconstruct the structures that are well seen in
Fig.~\ref{fig:model4eresid} (zoom-in panels).  The values of the
coefficients $\{T_i, a_i,\phi_i\}$ are provided as an
\textattachfile[author=Bartosz Lew,color=0 0 1,description=Fourier expansion coefficients table]{model5coeff.dat}{attachment}
to the on-line version of this article.

\begin{figure}
\centering
\includegraphics[width=0.49\textwidth]{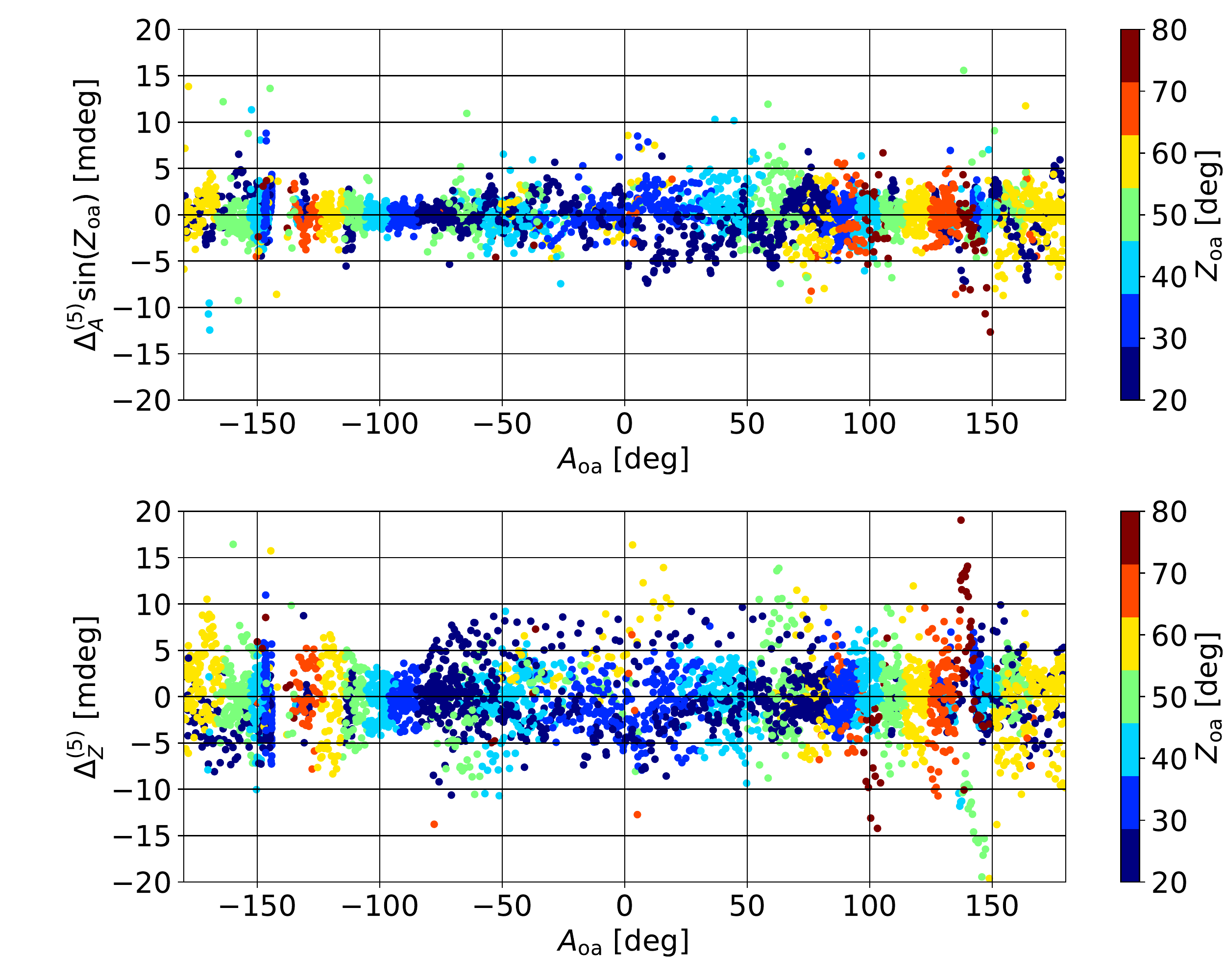}
\includegraphics[width=0.49\textwidth]{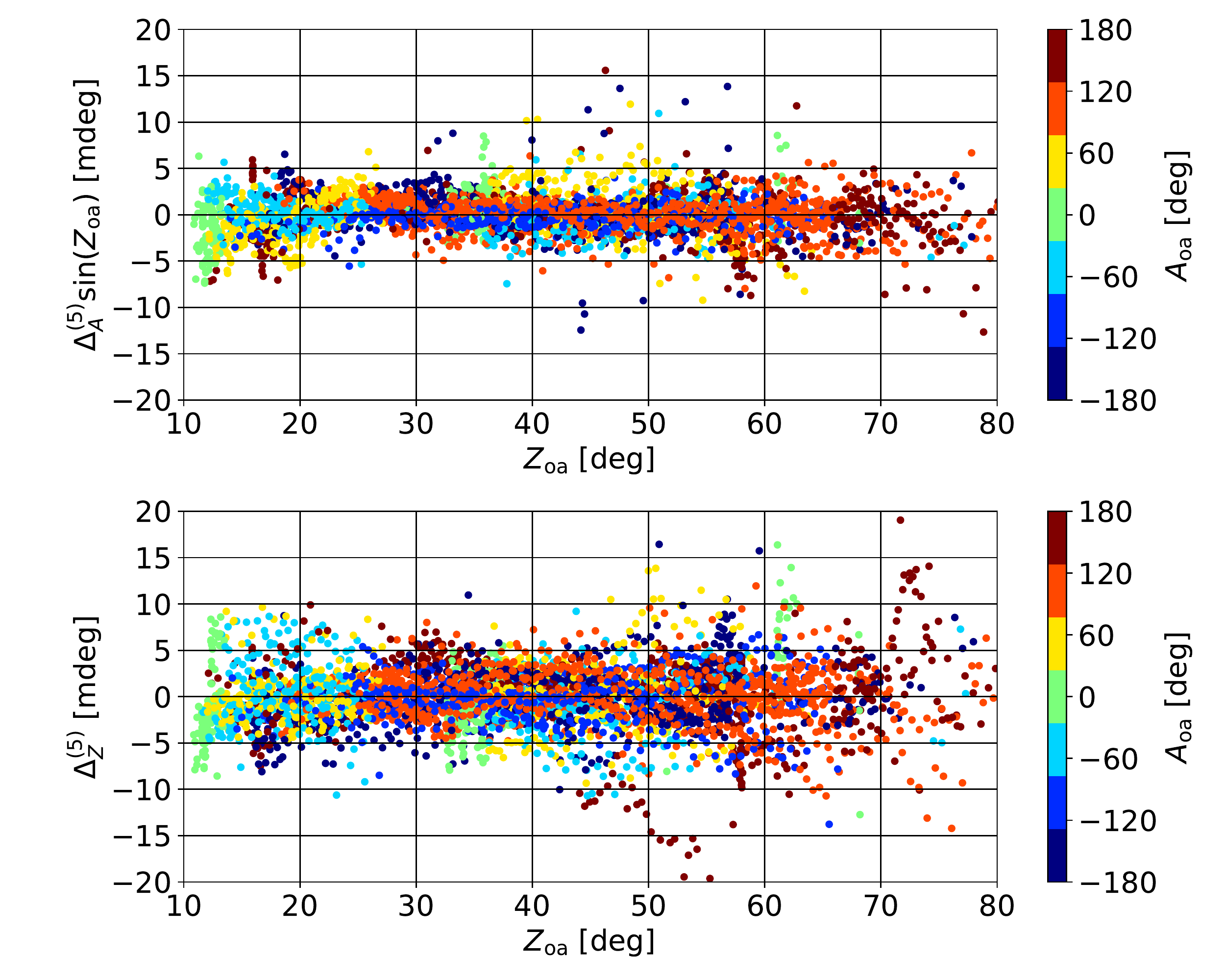}
\caption{Residuals between the pointing data and the best fit model
``Model 5'' ($\Model{5}$) plotted versus azimuth (upper
panels) and zenith distance (lower panels). The model effectively
removes all systematic effects associated with rail surface
irregularities (see Fig.~\ref{fig:model4efit}).}
\label{fig:model5resid}
\end{figure}

\begin{figure}
\centering
\includegraphics[width=0.4\textwidth]{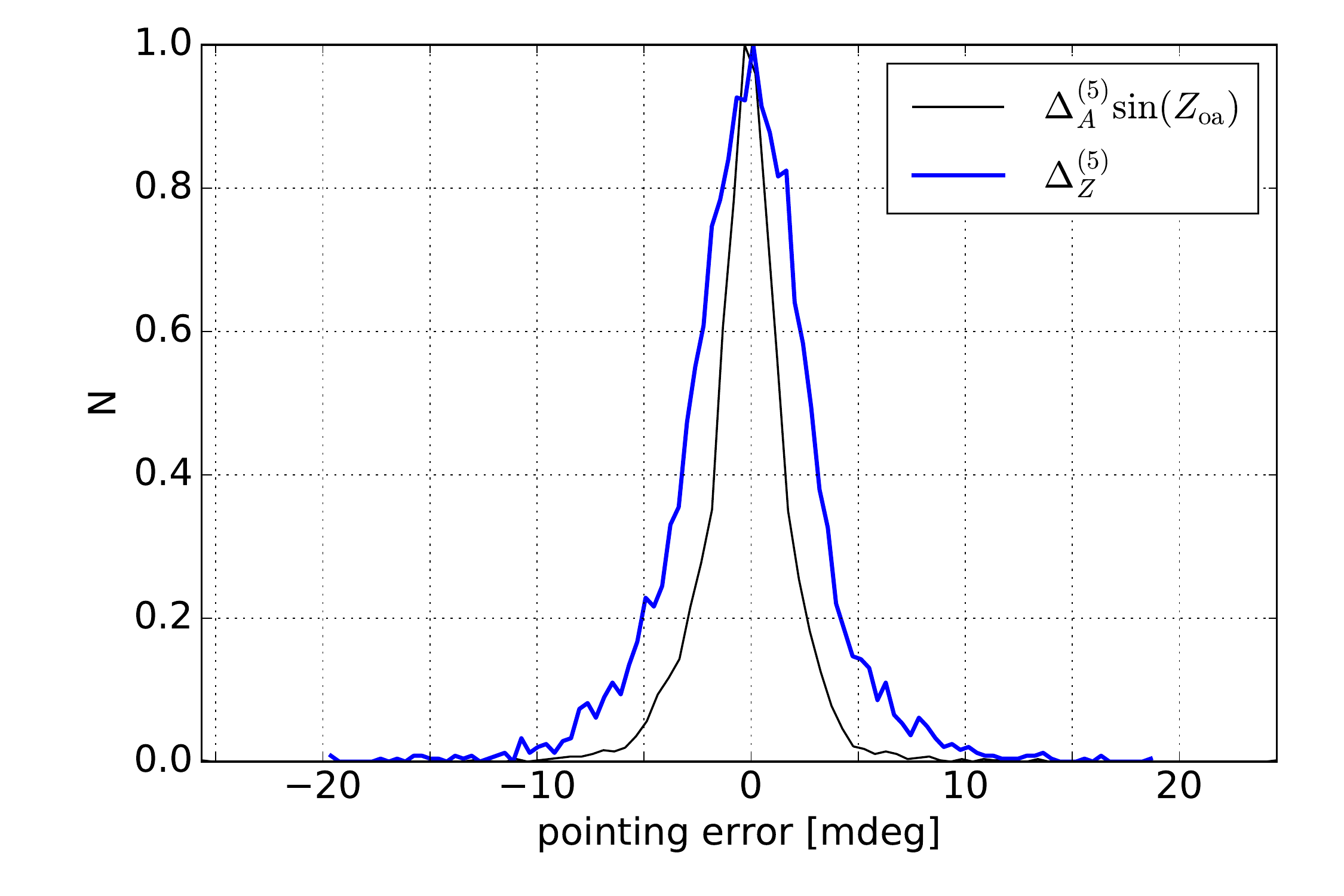}
\caption{Distribution of ``Model 5'' residuals from
Fig.~\ref{fig:model5resid}.  The RMS values of the residuals are
given in Table~\ref{tab:pointing}.}
\label{fig:model5residDistr}
\end{figure}

We introduce ``Model 5'' ($\Model{5}$) as:
\begin{subequations}
\begin{eqnarray}
\ModelA{5}({\bf p},\Aoa,\Zoa) &=& \ModelA{4e}({\bf p},\Aoa,\Zoa) + \delta_A(\Aoa)\,\,
\\
\ModelZ{5}({\bf q},\Aoa,\Zoa) &=& \ModelZ{4e}({\bf q},\Aoa,\Zoa) + \delta_Z(\Aoa)
\end{eqnarray}
\label{eq:model5} 
\end{subequations}
and we reprocess the corrections to obtain new residuals (Fig.~\ref{fig:model5resid}) and
their distribution (Fig.~\ref{fig:model5residDistr}).
In order to implement this model into a new version of the control system that we call {\coconut },
we use the approximation given in Eqs.~\ref{eq:approx}
since the control system calculates the corrections as a function of the
telescope encoder coordinates $(A_E, Z_E)$.

\begin{figure*}[!bht]
\centering
\includegraphics[width=\textwidth]{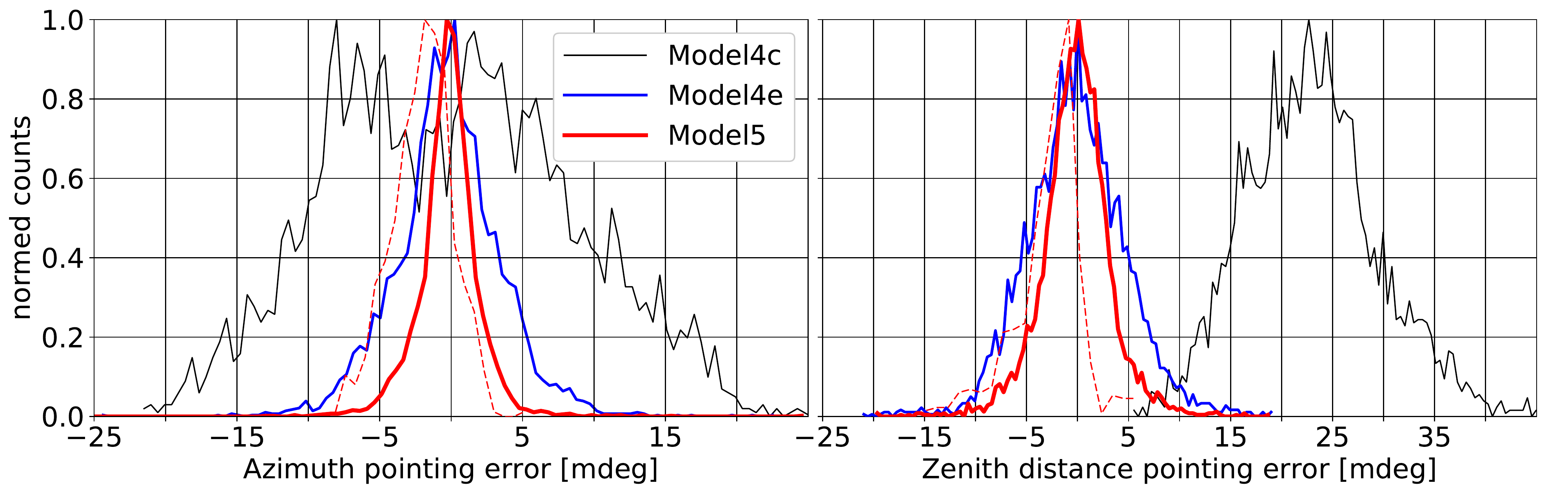}
\caption{Comparison of the
measured and anticipated pointing errors distributions, assembled
from Figs.~\ref{fig:fastTrackPointingError},
~\ref{fig:model4eresidDistr} and~\ref{fig:model5residDistr}, and
resulting from different pointing models. The azimuth pointing
errors are scaled by $\sin(\Zoa)$.  The RMS values for different
pointing models are given in Table~\ref{tab:pointing}. The dashed
line is a distribution of pointing corrections recorded after
``Model 5'' has been implemented in the telescope pointing system, and
it represents an additional validation of the model. A small,
${<}1.6/1.0$ mdeg systematic offset is present in azimuth/elevation,
witch can be associated with time-dependent effects that are being analyzed
and briefly discussed in Sec.\ref{sec:discussion}.}
\label{fig:modelsResidDistr} 
\end{figure*}

The large-scale fluctuations are clearly visible in ``Model 4e'' residuals
and they can be associated with the irregularities of the rail, that
is welded out of 14 pieces, which implies a period of
${\sim}25.7^\circ$. This period matches well the quasi-periodic
structures evident in Fig.~\ref{fig:model4eresid} (top-right) and
coincides with the number and the locations of the welding points
along the rail (Fig.~\ref{fig:model4efit}).  These saw-tooth--like
irregularities were previously detected in the pointing data analyzed
by \cite{Borkowski2006}, but the quality of that data was
substantially worse, and these effects have never been accounted for
in any pointing corrections model. A higher frequency modes of up to
${\sim}7^\circ$ deg are now also evident, and the current data allows
us to model them as well. The highest frequency modes that we model
also approximately correspond to the angular separation of the wheels
of the telescope trolleys.

In Fig.~\ref{fig:modelsResidDistr} we show a comparison between the
distributions of the measured position corrections in the {\fastTrack
} version of the control system and the anticipated improvements due
to using the new pointing models (``Model 4e'' and ``Model 5''). The
improvements are also shown by the RMS values in
Tab.~\ref{tab:pointing}.

\begin{table}[t]
\caption{Summary of measured and projected {\RT} pointing precision.}
\begin{tabular}{llllll}
\hline\hline
Control system & Pointing & RMS($A$) & RMS($Z$) & \multicolumn{2}{c}{Improvement\footnotemark[3]} \\
version & Model  &  [mdeg] &  [mdeg] & $A$& $Z$\\
\fastTrack & Model 4c &  8.5 & 24.1 & 1.0 & 1.0 \\
\coconut\footnotemark[1] & Model 4e\footnotemark[2] & 4.1 & 7.9 & 2.1 & 3.1\\
\coconut\footnotemark[1] & Model 4e & 3.7 & 4.8 & 2.3 & 5.0 \\
\coconut\footnotemark[1] & Model 5 & 2.2 & 3.4 & 3.9 & 7.1 \\
\coconut\footnotemark[4] & Model 5 & 2.7 (2.1) & 3.9 (3.2) & 3.1 & 6.2 \\
\end{tabular}

\footnotetext[1]{Anticipated precision for the pointing model.}
\footnotetext[2]{The best fit assuming $\xi_A=\xi_Z=\xi$ and
$\zeta_A=\zeta_Z=\zeta$ in Eqs.~\ref{eq:model4e}}
\footnotetext[3]{RMS improvement with respect to the {\fastTrack} version of the control system.}
\footnotetext[4]{Precision measured using an additional pointing data taken after the pointing model
was implemented into the control system. Standard deviations are also
given in parentheses for comparison, in order to indicate yet
unidentified, long-term systematic effects that are still not accounted for in the model
(see Fig.~\ref{fig:modelsResidDistr} and Sec.~\ref{sec:discussion}
for a discussion).}
\label{tab:pointing}
\end{table}

\section{Measurements of rail and wheels irregularity}
\label{sec:rail_and_wheel}
If the telescope wheels are not exactly round, but rather oval to the
first approximation, then the fast running pointing corrections may
only partially be associated with the rail irregularities as the
wheels of 1.4~m in diameter cover azimuth range of about $21^\circ$
per single rotation.\footnote{The highest frequency Fourier modes that
model the rapidly changing position corrections have period of about
$7^\circ$ (Sec.~\ref{sec:model5}).}  Furthermore, the time stability
of the rail model (Fig.~\ref{fig:model4efit}, Eq.~\ref{eq:model5fit})
would be uncertain if e.g. wheels could slip over the rail, or roll
along non-repeatable paths. In fact, we observed that matching a pair
of points, one on the rim of a given wheel and another one on the
rail, and both fixed at the same azimuth, is not stable in time.

\begin{figure*}
\centering
\includegraphics[width=.98\textwidth]{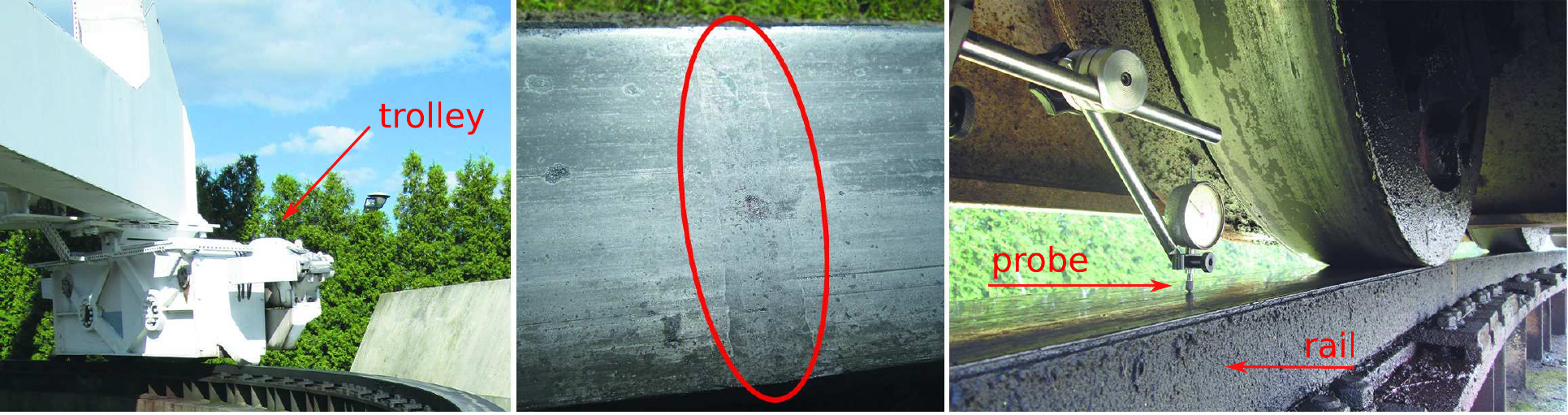}
\caption{Rail surface irregularity measurement.
The response of the rail to the pressure induced by two-wheeled
trolleys supporting the telescope structure ({\em left}) is measured
for all azimuths using readings from an analog distance indicator
attached to the trolley chassis ({\em right}). The largest variations
are observed in the vicinity of rail welding points ({\em center}).}
\label{fig:rail_meas}
\end{figure*}

Therefore, we measure the wheel and rail irregularities using an
analog distance indicator, which offers a relative distance
measurement accuracy of $0.01$~mm.  In each case the distance
indicator is attached to the trolley chassis
(Fig.~\ref{fig:rail_meas}).  For the case of wheels, the indicator
probe is set against the wheel surface and points at its center.  As
the telescope moves a video camera records the probe indications and
the recording is repeated for each wheel. For the case of the rail,
the probe is fixed vertically against the rail surface and the
recording is done for two dish orientations: $Z=0^\circ$ and
$Z=60^\circ$.  We measure the rail deformations with the probe
installed at two different trolleys: front and back, at two different
distances from the wheel: ${\sim}60$ cm and ${\sim}140$ cm
respectively.

The azimuthal extent of rail deformations is unknown, and in our
measurement setting (Fig.~\ref{fig:rail_meas}) the indicator readings
depend on the distance of the probe from the wheel.  For the case when
the probe is fixed nearby the wheel, we anticipate that the measured
value of the vertical rail deformation ($\Delta h$) is biased low,
since in this case the probe will capture only a fraction of the full
indent.  In the extreme case, if the probe was set to measure the
distance variations (from the chassis to the rail) right under the
wheel, no variation would be detected (assuming that wheels are
round).  The distance of 60 cm from the wheel is about a factor of 2
smaller than the distance that corresponds to the azimuth range
(${\sim}7^\circ$) over which we observe significant variations in the
measured position corrections, which we associate with the rail
irregularity (Fig.~\ref{fig:model4efit}).  Therefore, we also measure
the track height variations with the probe fixed about 1.4 m away from
the wheel. At this distance we expect to capture most of the rail
indents.

An analysis of the relative variations of the wheels radii, shows that
the maximal deviation from the circular shape is ${<}0.05$~mm.  This
is negligible when compared to the amplitude of the rail indents
around the welding points: ${<}2.7/4.4$~mm with the probe fixed at the
distance of about $60/140$ cm from the wheel in one of the front/back
trolleys.
We find that the dependence of these values on the dish elevation is
weak (a difference of ${<}0.2$ mm between $Z=0^\circ$ and
$Z=60^\circ$).

Assuming that a rail indent of ${\approx}4.4$~mm can generate a
pointing offset of up to ${\sim} 20$ mdeg (Fig.~\ref{fig:model4efit}),
it should be expected that the effects due to non-round wheels will
limit the pointing accuracy to ${<}0.25$ mdeg.
This is below the pointing precision we aim at, and therefore the
effects due to non-round wheels can be neglected at the present.

\subsection{Impact on pointing corrections}
\label{sec:rail_impact}
It is instructive to consider a toy model of rail surface
instability in order to estimate the expected amplitude of the
associated pointing corrections.

The radius of the rail is $R=12\,{\rm m}$, and the four trolleys
supporting the telescope structure are located at corners of a
square frame (Fig.~\ref{fig:rail_meas}).
Assuming that the telescope structure is rigid and that one of the
trolleys is rolling down a local rail indent of the amplitude
$\Delta h$, the opposite trolley will be lifted (or will be less
loaded)
and the vertical axis of the telescope will be rotated about the
axis defined by the two remaining trolleys. The rotation angle will
be:
\begin{equation}
\gamma =\arcsin\left(\frac{\Delta h}{R}\right) \approx 4.8 \left(\frac{\Delta h}{1\,{\rm mm}}\right)\,{\rm [mdeg]}
\label{eq:gamma}
\end{equation}
Assuming that the largest amplitude of the rail indents (registered
over the full azimuth rotation) is $\Delta h{\approx}4.4$ mm, the
expected $\gamma$ is about $21$ mdeg.  It is straightforward to
calculate (App.~\ref{sec:rail_and_pointing}) that, in the limit of
small angles, the associated pointing corrections are:
\begin{subequations}
\label{eq:rail_indent}
\begin{eqnarray}
\Delta A \sin(Z) &=& \gamma \cos(Z)/\sqrt{2}\\
\Delta Z &=& \gamma/\sqrt{2},
\end{eqnarray}
\end{subequations}
which gives the expected amplitude of the corrections at the
level ${<}15$ mdeg in both coordinates.

In Fig.~\ref{fig:model4efit} the amplitude of the systematic effects,
that we associate with rail surface irregularities, ranges from about
16 mdeg to 18 mdeg for elevation and cross-elevation respectively.

Since these values are within a factor of ${\sim}1.2$ from the values
predicted by the model at the zenith, it is clear
that the toy model, given a realistic rail indent amplitude measurement,
predicts the pointing corrections that are quite
consistent with the observations discussed in the previous sections
(Sec.~\ref{sec:model5}). However, the effects of rail
indents are operating in all four trolleys simultaneously and possibly
are also associated with some structural deformations, therefore in this work
we model the position corrections due to the rail instability relying
solely on pointing measurements.
\section{Discussion}
\label{sec:discussion}
Although ``Model 5'' predicts an improvement in pointing precision by a
factor of a few with respect to the {\fastTrack } version of the
control system (Tab.~\ref{tab:pointing}), Fig.~\ref{fig:model5resid}
also hints that the pointing data, when corrected for the best-fit
pointing model ($\Model{5}$), may still have some residual zenith
distance dependence. For example, in the azimuth range
$A=[50^\circ,70^\circ]$ the azimuth residuals seem to be smaller at
lower elevations than the residuals of the measurements taken at higher
elevations.  The same seems to be true for the zenith distance residuals in
the azimuth range $A=[-180^\circ,-150^\circ]$.
Whether this is a coincidence is not clear at the present.
The azimuth and zenith distance residua (Fig.~\ref{fig:model5resid})
clearly depend on elevation around $A=140^\circ$. New pointing data
may help to better understand the remaining issues around those and
other directions.

The data presented in this paper are the most accurate pointing
observations ever collected with {\RT} (Fig.~\ref{fig:model4eresid}
right panels).  However, the noise level of the azimuth residuals is
not uniform across the full range of azimuths
(Fig.~\ref{fig:model5resid}).  Generally, the azimuth residuals have
smaller dispersion than the zenith distance residuals. It remains to be
seen whether the larger dispersion in zenith distance can be
associated with imperfections in fixation of the secondary
mirror. Taking into account results from independent observations of
the position of the secondary mirror, which are performed using an optical
camera installed in the secondary focus cabin, this possibility seems plausible.
However, by inspecting the residuals in Fig.~\ref{fig:model5resid}, it
is clear that at the current noise level, no obvious nor strong
large-scale systematical effects are present, which suggests that the
extended pointing model (Eqs.~\ref{eq:model5}) accounts for all major
construction deficiencies.

In principle, the dimensionality of the model parameter space
could be reduced by utilizing independent measurements
of the gravitational sag, that can be obtained with aid of the
large-zoom optical camera, installed in the secondary focus cabin,
which we use for real-time monitoring of the position of the secondary mirror.
However, the gravitational sag can be a function of temperature
and it is possible that a general solution may require a more complicated
model of the sag. We will investigate this possibility in another study
(\citeLewInPrep).

The pointing models considered in this work do not accommodate for
horizontal focus box offsets that depend on elevation, nor for
non-vanishing azimuth dependence of the gravitational sag. Results
from optical imaging suggest that such effects may be present, but 
they are not dominant.  The anticipated pointing precision
of the {\coconut } version of the control system (${<}3.4$ mdeg) is
still at least a factor of a few below the tracking capability of the
telescope. Yet higher accuracy pointing data would be needed to
explore these possibilities in greater details and possibly further
improve the pointing.

The data analyzed in the this paper (Sec.~\ref{sec:data}) do not
provide multiple observations of same directions, and the sky coverage
is incomplete. With the advent of new pointing data, to be taken with
the {\coconut } version of the control system (utilizing ``Model 5''),
it will be possible to verify and investigate the time stability of
the corrections, which is of crucial importance for any pointing model
that is calibrated against observational data taken over a short time
interval. This will also address the important issues of short- and
long-term thermal effects on the pointing performance as the amplitude
of annual temperature variations at the telescope site typically spans
well over $40^\circ$C.

The ``Model 5'' has been implemented into control system in December
2016 and has been tested since then.  The actual pointing measurements
taken right after the implementation of the model (but a few months
after the acquisition of the data used to calibrate the models
presented in this work) confirm the anticipated pointing precision
(Tab.~\ref{tab:pointing}). However, when combined with large
zoom video camera observations of the secondary mirror position,
they also hint at the presence of new long- and short-term thermal or
time-dependent systematic effects, which are being analyzed
(Fig.~\ref{fig:modelsResidDistr}). For example, if during long
integrations the anisotropic illumination by the Sun could induce a
temperature difference of $\Delta T=5^\circ$C between the supporting
legs, then the corresponding difference in leg lengths of ${\sim}1$ mm
should be expected due to thermal expansion.\footnote{We assume the
effective length of the legs of $L\approx18$~m, and the linear
thermal expansion coefficient for iron $\alpha=12\times
10^{-6}\,{\rm K^{-1}}$.}  These effects could translate onto
pointing corrections of the amplitude similar to those caused by the
rail surface irregularities.  Some of the new pointing data
also show fast varying, then disappearing pointing abnormalities,
but those are typically excluded at the data selection stage.
Similar effects have also been previously noted in \cite{Bayley1994}
for a comparable telescope.  In such a case, accounting for thermal
effects could be important for reaching stable, milli-degree
pointing, but this exceeds beyond the scope of the this work.

\section{Conclusions}
\label{sec:conclusions}
We implement a number of astrometric improvements into the control
system of the 32-meter radio telescope located near Toru\'n.
Using a dedicated software toolkit we acquire pointing data from
observations carried out in 2016, and we analyze them in order to
improve the pointing precision.

We slightly modify and extend the pointing corrections model used
during the data taking, and we refer to it as ``Model 4e''. Using the
pointing data and $\chi^2$ minimization approach, we find the values
of 16 parameters of the model that minimize the pointing correction
residuals.  The best-fit model can improve the pointing precision to
$13''$ ($17''$) in azimuth (elevation) as measured by the
RMS of the residual pointing corrections, or by a factor of $2.3$ ($5.0$)
with respect to the pointing precision available at
the time of the data taking.

Next, we analyze the pointing corrections processed with
``Model 4e''. We find systematic fluctuations in the residuals with
peak-to-peak amplitude of up to ${\sim}20$ mdeg, which we identify to be
associated with the telescope rail. The rail yields under the weight of
620-ton telescope as four two-wheeled supporting trolleys roll over
and around the rail welding points.  The pointing effects associated
with the rail vertical deformations, which amplitude we measure to be
${<}4.4$~mm, cannot be modeled with ``Model 4e''.  In order to account
for these irregularities we introduce an extension to the model that
we call ``Model 5''. The extension allows us to further improve the
pointing accuracy down to $8''$ ($12''$) in azimuth (elevation) or by
a factor $1.7$ ($1.4$) with respect to the best fit model
(``Model 4e'') that does not account for the rail irregularities.

We also estimate that deviations of telescope wheels from circular shape of
the amplitude ${<}0.05$~mm can effectively limit the pointing accuracy
to ${<}0.25$ mdeg, which is small when compared to other effects
investigated in this work.  The upper limit of the deviations of the
wheels from circular shapes indicates that the most rapidly
changing pointing corrections associated with the rail should be
stable over time even if the wheels slip.

The pointing corrections when processed through ``Model 5'' show no
evident systematic effects in residuals, suggesting that at the current
noise level the model accounts for all major effects that contribute
to pointing errors. However, we note that the dispersion of zenith
distance corrections is larger than it is in the case of azimuth
corrections, which may caused by instabilities of the suspension of
the secondary mirror or thermal effects.

The data used for fitting the pointing model were acquired within a
relatively short period of time -- about 3.5 months of the summer
time.  While this is may be enough to fit a model, any systematic long
term effects (associated with e.g. seasonal temperature variations),
if present, may be missed. However, we have shown that if thermal or time
dependent effects are under control, the telescope should be capable of pointing with
${<}12''$ accuracy.

\section*{Acknowledgments}
BL would like to thank Kaz Borkowski for his valuable comments on the
manuscript and for his previous works on {\RT} pointing models.
BL also would like to thank Jacek Kr\'ol, Wojtek Szyma\'nski, Jacek
Jopczy\'nski, Janusz Filarecki, Gieniu Pazderski and Pawe\l{} Wolak
for the maintenance of the telescope secondary mirror and the initial
measurements of wheels and the rail.
Also, thank you to Marcin Gawro\'nski for help in finding suitable
position calibrators, to Marian Szymczak for comments and suggestions,
and to Boud Roukema for reading the manuscript.
BL wishes to thank anonymous referee for stimulating comments
that helped improving the manuscript and gaining new insights.

We acknowledge use of the 'matplotlib' plotting library \citep{Hunter2007}.

This work is based in part on observations carried out using the
32-meter radio telescope operated by Toruń Centre for Astronomy of
Nicolaus Copernicus University in Toru\'n (Poland) and supported by the
Polish Ministry of Science and Higher Education SpUB grant.

\appendix

\section{Model of pointing corrections due to rail height irregularities}
\label{sec:rail_and_pointing}
\begin{figure}
\centering
\includegraphics[width=0.35\textwidth]{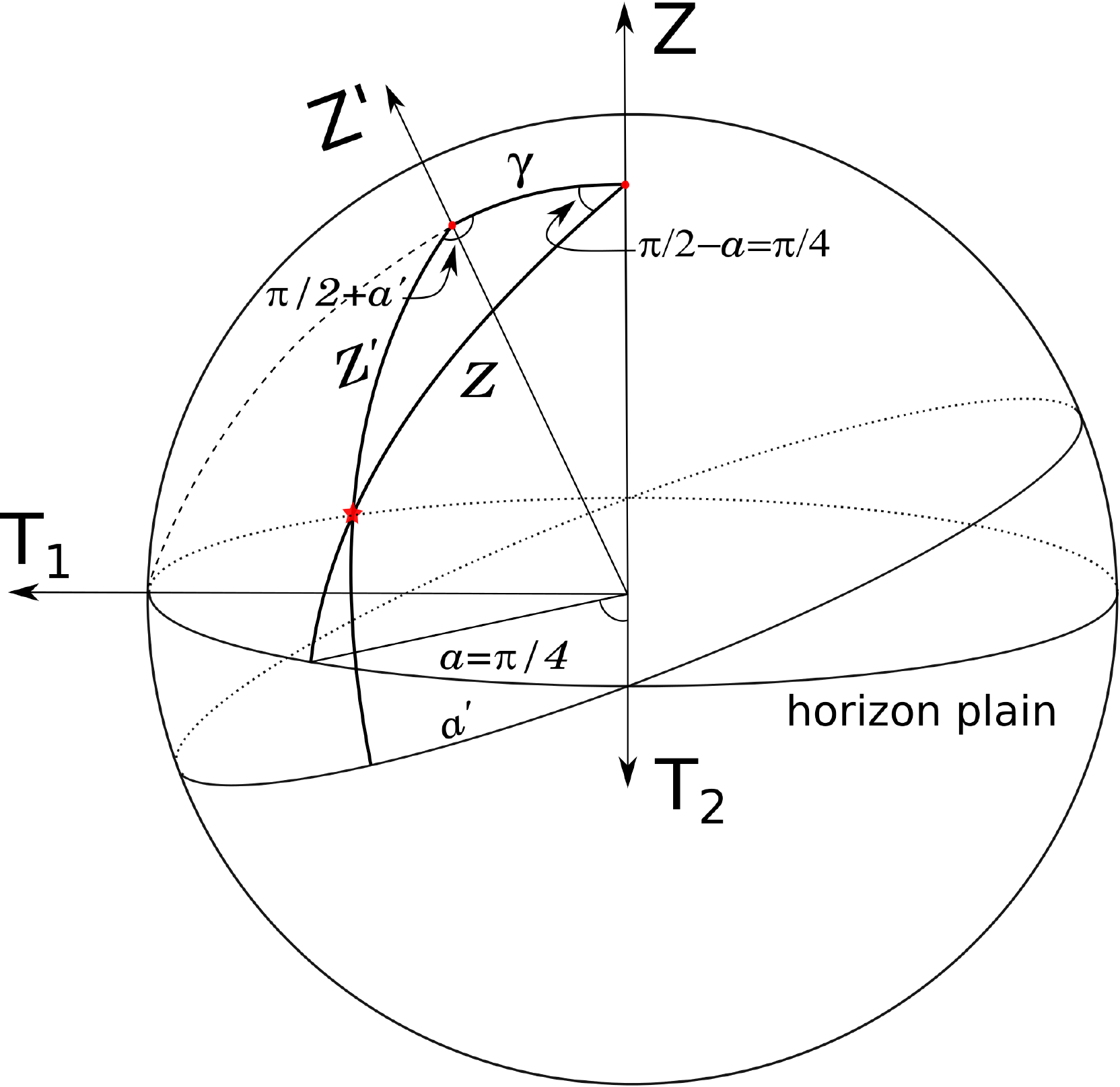}
\caption{A model of the effects of rail indent on telescope structure tilt
(see text for comments).
}
\label{fig:sphere}
\end{figure}

In this section we calculate the azimuth and zenith distance pointing
corrections due to telescope tilting caused by a local rail indent.
We use a simple geometrical model, previously considered by Kaz
Borkowski.
We assume that the four trolleys
of the telescope are connected by a rigid square frame
(Sec.~\ref{sec:rail_impact}). The telescope weight split between the
trolleys is, in general, not even and may depend on the dish
elevation. For example, counter-weights heavier than the dish will
cause larger loading of back trolleys with respect to the front ones.
When one of the heavily loaded trolleys rolls down an indented region
of the rail, the telescope zenith axis will tilt towards that trolley,
thus altering the source apparent zenith distance and azimuth. In
Fig.~\ref{fig:sphere}, the Z axis represents the geodetic zenith of a
perfect telescope, whereas Z' represents a zenith tilted by an angle
$\gamma$ towards the trolley located on axis ${\rm T_{1}}$. The points
where axes Z and Z' cross the celestial sphere are marked with red
dots.  The axis in the horizon plain, ${\rm T_{2}}$, which is
perpendicular to ${\rm T_{1}}$ is the rotation axis. Without indents,
these two axes co-rotate with the telescope in azimuthal motion. The
relative azimuth, $a$, of the telescope pointed at a source (marked
with a star) is calculated from axis ${\rm T_2}$ and, by construction,
is always equal to $\pi/4$ (or $\pi/4+\pi$ depending on which side the
telescope zenith is tilted). The source is found at the zenith distance $Z$ from
axis Z, and at the zenith distance $Z'$ from axis Z'.

The spherical triangle with vertices at the geodetic
zenith (Z), the tilted zenith (Z') and at the source (Fig.~\ref{fig:sphere})
yields:
\begin{subequations}
\label{eq:sphere}
\begin{align}
\sin Z'\sin\left(\frac{\pi}{2}+a'\right)&=\sin Z\sin\frac{\pi}{4}\\
\sin Z'\cos\left(\frac{\pi}{2}+a'\right)&=\sin\gamma\cos Z-\cos\gamma\sin Z\cos\frac{\pi}{4}\\
\cos Z'&=\cos\gamma\cos Z + \sin\gamma\sin Z\cos\frac{\pi}{4},
\end{align}
\end{subequations}
and it is straightforward to calculate the
expected pointing corrections in azimuth and zenith distance:
\begin{subequations}
\label{eq:rail_indent2}
\begin{align}
\tan a'-\tan a &=\cos\gamma - 1 - \sqrt{2}\sin\gamma\cot Z\\
\cos Z'-\cos Z &=(\cos\gamma-1)\cos Z+\frac{\sqrt{2}}{2}\sin\gamma\sin Z,
\end{align}
\end{subequations}
which in the limit of small angles become:
\begin{subequations}
\label{eq:rail_indent3}
\begin{align}
\Delta A \sin(Z) &= (a-a')\sin(Z) \approx \gamma\cos Z/\sqrt{2}\\
\Delta Z &= Z-Z' \approx \gamma/\sqrt{2}.
\end{align}
\end{subequations}

\bibliography{bibliography} 
\bibliographystyle{aaads}

\end{document}